\documentclass[longauth]{aa} 

\usepackage{graphicx}
\usepackage{graphicx}	
\usepackage{amsmath}	
\usepackage{amssymb}	
\usepackage{caption}
\usepackage{lscape}
\usepackage{verbatim}
\usepackage{hyperref}
\usepackage{txfonts,xcolor}
\usepackage{rotating}
\usepackage{lineno}
\usepackage{pdflscape}
\usepackage[colorinlistoftodos]{todonotes}
\def\arcmin{\hbox{$^\prime$}}
\def\arcsec{\hbox{$^{\prime\prime}$}}

\usepackage{scalerel}
\usepackage{tikz}
\usetikzlibrary{svg.path}
\definecolor{orcidlogocol}{HTML}{A6CE39}
\tikzset{
  orcidlogo/.pic={
    \fill[orcidlogocol] svg{M256,128c0,70.7-57.3,128-128,128C57.3,256,0,198.7,0,128C0,57.3,57.3,0,128,0C198.7,0,256,57.3,256,128z};
    \fill[white] svg{M86.3,186.2H70.9V79.1h15.4v48.4V186.2z}
                 svg{M108.9,79.1h41.6c39.6,0,57,28.3,57,53.6c0,27.5-21.5,53.6-56.8,53.6h-41.8V79.1z M124.3,172.4h24.5c34.9,0,42.9-26.5,42.9-39.7c0-21.5-13.7-39.7-43.7-39.7h-23.7V172.4z}
                 svg{M88.7,56.8c0,5.5-4.5,10.1-10.1,10.1c-5.6,0-10.1-4.6-10.1-10.1c0-5.6,4.5-10.1,10.1-10.1C84.2,46.7,88.7,51.3,88.7,56.8z};
  }
}

\begin{document}

\title{Multi-wavelength Constraints on the Transient EP250905a}


\author{
J. Quirola-V\'asquez\inst{1}
\and
P.~G. Jonker\inst{1}
\and
A. Levan\inst{1}
\and
D. B. Malesani\inst{2,3}
\and
F. E. Bauer\inst{4}
\and
A. Martin-Carrillo\inst{5}
\and
\hbox{G. Corcoran}\inst{5}
\and
D. Mata S\'anchez\inst{6,7}
\and
R. A. J. Eyles-Ferris\inst{8}
\and
F. Carotenuto\inst{9}
\and
M. Ravasio\inst{10,11}
\and
\hbox{J. S\'anchez-Sierras}\inst{1}
\and
J. Bright\inst{12}
\and
J. A. Chac\'on\inst{13}
\and
L. Cotter\inst{5}
\and
F. J. Cowie\inst{14}
\and
N. Sarin\inst{15,16}
\and
\hbox{M. A. P. Torres}\inst{6,7}
\and
J. N. D. van Dalen\inst{1}
\and
A. P. C. van Hoof\inst{1}
\and
V. D’Elia\inst{17}
\and
P. Jakobsson\inst{18}
\and
N. Habeeb\inst{19}
\and
\hbox{S. Kobayashi}\inst{20}
\and
A. Saccardi\inst{21,22}
\and
M. De Pasquale\inst{23}
\and
D. Xu\inst{24}
\and
Y. H. Cheng\inst{25}
\and
R. D. Liang\inst{24,26}
\and
H. Sun\inst{24}
\and
Y. Wang\inst{27,28}
\and
\hbox{W. Yuan}\inst{24,29}
\and
W. D. Zhang\inst{24}
}

\institute{
Department of Astrophysics/IMAPP, Radboud University, P.O. Box 9010, 6500 GL, Nijmegen, The Netherlands
\and
The Cosmic Dawn Centre (DAWN), Denmark
\and
Niels Bohr Institute, University of Copenhagen, Jagtvej 155, DK-2200, Copenhagen N, Denmark
\and
Instituto de Alta Investigaci{\'{o}}n, Universidad de Tarapac{\'{a}}, Casilla 7D, Arica, Chile
\and
School of Physics and Centre for Space Research, University College Dublin, Belfield, Dublin 4, Ireland
\and
Instituto de Astrof\'isica de Canarias, E-38205, La Laguna, Tenerife, Spain
\and
Universidad de La Laguna, Departamento de Astrof\'isica, E-38206, La Laguna, Tenerife, Spain
\and
School of Physics and Astronomy, University of Leicester, University Road, Leicester, LE1 7RH, UK
\and
INAF – Osservatorio Astronomico di Roma, Via Frascati 33, I-00078 Monte Porzio Catone (RM), Italy
\and
Institute of Space Sciences (ICE, CSIC), Campus UAB, Carrer de Can Magrans s/n, E-08193, Barcelona, Spain
\and
Institut d'Estudis Espacials de Catalunya (IEEC), 08860 Castelldefels (Barcelona), Spain
\and
Astrophysics, Department of Physics, The University of Oxford, Keble Road, Oxford, OX1 3RH, UK
\and
Instituto de Astrof{\'{i}}sica, Facultad de F{\'{i}}sica, Pontificia Universidad Cat{\'{o}}lica de Chile, Campus San Joaqu{\'{i}}n, Av. Vicuña Mackenna 4860, Macul Santiago, Chile, 7820436
\and
Department of Physics, University of Oxford, Denys Wilkinson Building, Keble Road, Oxford OX1 3RH, UK
\and
Kavli Institute for Cosmology, University of Cambridge, Madingley Road, CB3 0HA Cambridge, UK
\and
Institute of Astronomy, University of Cambridge, Madingley Road, CB3 0HA, UK
\and
Space Science Data Center (SSDC) – Agenzia Spaziale Italiana (ASI), I-00133 Roma, Italy
\and
Centre for Astrophysics and Cosmology, Science Institute, University of Iceland, Dunhagi 5, 107 Reykjavík, Iceland
\and
School of Physics and Astronomy, University of Leicester, University Road, Leicester LE1 7RH, UK
\and
Astrophysics Research Institute, Liverpool John Moores University 146 Brownlow Hill, Liverpool L3 5RF, UK
\and
Université Paris-Saclay, Université Paris Cité, CEA, CNRS, AIM, 91191 Gif-sur-Yvette, France
\and
Centre national d’études spatiales (CNES), Paris, France
\and
University of Messina, Mathematics, Informatics, Physics and Earth Science Department, Via F.S. D’Alcontres 31, Polo Papardo, 98166 Messina, Italy
\and
National Astronomical Observatories, Chinese Academy of Sciences, Beijing 100101, China
\and
South-Western Institute for Astronomy Research, Yunnan University, Kunming 650504, China
\and
University of Chinese Academy of Sciences, Chinese Academy of Sciences, Beijing 100049, China
\and
Purple Mountain Observatory, Chinese Academy of Sciences, Nanjing 210023, China
\and
Department of Astronomy, University of California, Berkeley, CA 94720-3411, USA
\and
School of Astronomy and Space Sciences, University of Chinese Academy of Sciences, Beijing, 100049, People’s Republic of China
}

\abstract
{Fast X-ray transients (FXTs) are a diverse class of high-energy suggested origins, ranging from stellar explosions to compact object mergers. The Einstein Probe (EP) satellite discovers approximately 100 FXTs per year.}
{We aim to constrain the physical origin of EP250905a.}
{We analyze X-ray, optical, near-infrared (NIR), and radio temporal and spectral properties of EP250905a. In addition, we assess the possible role of weak gravitational lensing in shaping its observed characteristics.}
{EP250905a fades rapidly in X-rays, and we detect no NIR or radio emission, but we detect early optical emission that rapidly fades beyond the detection limits. Two nearby galaxies are identified for which we derive spectroscopic redshifts of $z=0.374$ (G1) and $z=2.714$ (G2). Our analysis favors G2 as the host of the FXT EP250905a. The angular separation of 2.56\arcsec\, between the FXT's optical counterpart and the center of the G1 galaxy suggests the emission of the FXT might be moderately magnified by lensing effects ($\mu\approx3.9$) given the inferred Einstein radius of G1 ($\theta_E\approx1.9\arcsec$).}
{The data are best explained as an afterglow from a mildly off-axis structured jet at $z=2.714$, providing a consistent broadband interpretation that also allows for weakly lensed emission of EP250905a.}

\keywords{X-rays: general – X-rays: bursts – Gamma-ray bursts}
\maketitle
\nolinenumbers

\section{Introduction}

Extragalactic fast X-ray transients (FXTs) are defined as bursts of soft X-ray photons (in the ${\sim}$0.3--10~keV band) with durations from seconds to hours \citep{Heise2010}.
Over the last two decades attempts have been made to systematically identify and characterize sources detected by the new generation of X-ray satellites such as the Neil Gehrels \emph{Swift} Observatory \citep[\emph{Swift}; e.g.,][]{Soderberg2008,Evans2023}, the \emph{Chandra} X-ray Observatory \citep[e.g.,][]{Jonker2013,Glennie2015,Irwin2016,Bauer2017,Xue2019,Lin2022,Quirola2022,Quirola2023}, and the \emph{X-ray Multi-mirror Mission - Newton telescope} \citep[\emph{XMM-Newton}; e.g.,][]{Novara2020,Alp2020,DeLuca2021}.
The high spatial resolution of these observations has robustly established the population of extragalactic X-ray transients, in addition to X-ray outbursts from Galactic systems, i.e., strengthening their extragalactic nature and even enabling the localization of the host galaxies of the transients \citep[e.g.,][]{Lin2022,Eappachen2022,Eappachen2023a,Eappachen2024,Quirola2024b}.

The launch of the \emph{Einstein Probe} \citep[EP;][]{Yuan2015,Yuan2022,Yuan2025} satellite on January 9, 2024, is enabling a transformation of the field, discovering $\approx$220 new extragalactic FXTs in the first two years of operations.
EP has two instruments onboard. The first instrument is the Wide-field X-ray Telescope (EP-WXT), an imaging survey instrument operating in the soft X-ray band (0.5--4 keV) with a field of view (FoV) of 3600 square degrees. It is ideally suited for the arcminute localization and rapid reporting of FXTs.
The second instrument onboard is the EP Follow-up X-ray Telescope (EP-FXT). EP-FXT often provides an X-ray position accurate to a few tens of arcseconds on timescales of orders of an hour or even less when the EP-WXT triggers an automatic slew to the EP-FXT instrument. The rapid follow-up enabled by EP has led to the detection of the optical, near-infrared (NIR), and radio counterpart to a subsample of tens of FXTs \citep[e.g., EP240315a, EP240414a, EP240806a, EP250108a, and EP250304a;][]{EP240315a_GCN,EP240414a_GCN,EP240806a_GCN,Gillanders2024,EP250108a_GCN,EP250304a_GCN,Bright2025}.

A variety of progenitor scenarios have been proposed to explain the X-ray light curves and spectra of extragalactic FXTs, as well as the expected multi-wavelength emission of their counterparts. These include:
\begin{itemize}
    \item Shock breakout (SBO) associated with certain types of core-collapse supernovae (CC-SNe), which can produce an FXT when the shock wave generated during the explosion reaches the surface of the progenitor star \citep[e.g., SN2008D/XRF~080109 and in the last months EP260321a;][]{Soderberg2008,Waxman2017,Huang2026_GCN,Rastinejad2026,OConnor2026,Wen2026,Chen2026,Yuan2026,Martin2026};
    
    \item Models involving binary neutron star (BNS) mergers that lead to the formation of a rapidly rotating, highly magnetized neutron star (magnetar), capable of powering a nearly isotropic X-ray signal observable as an FXT \citep[e.g., XRT~141001, XRT~150322, and EP~250207b;][]{Dai2006,Metzger2008,Yu2013,Zhang2013,Sun2017,Sun2019,Quirola2024,Jonker2025}, although see van Hoof et al. in prep. for the case of EP250207b (where an supernova cannot be ruled out); 
    
    \item A certain connection between some FXTs and broad-lined Type Ic SNe \citep[e.g., EP240414a, EP240801a, and EP250108a;][]{Rastinejad2025,Eyles_Ferris2025,Srinivasaragavan2025,Li2025b,van_Hoof2026};
    
    \item A possible link of FXTs with luminous fast blue optical transients \citep[LFBOTs; e.g., EP240414a;][]{van_Dalen2024,Srivastav2025,Sun2025};

    \item High-redshift, intrinsically low-luminosity, and seen off-axis gamma-ray bursts (GRBs), whose emission is shifted into the X-ray band due to cosmological redshift or intrinsic geometry \citep[e.g., EP240315a, EP250321a and EP260119a;][]{Wichern2024,Levan2024,Gillanders2024,Zhu2025_GCN,Corcoran2026_GCN}; and
    
    \item Tidal disruption events (TDEs), particularly those involving a white dwarf disrupted by an intermediate-mass black hole (${\sim}10^3$--$10^5$~M$_\odot$), which may give rise to FXT-like emission \citep[e.g., XRT000519, EP240408a, and EP250702a;][]
{Jonker2013,MacLeod2016,Maguire2020,OConnor2025,Levan2025,Eyles-Ferris2026,Li2026,Gompertz2026}.
\end{itemize}

The X-ray transient EP250905a was detected by the EP-WXT instrument 
on 2025 September 5, at coordinates RA$_{\rm J2000} = 4.441^\circ$, Dec$_{\rm J2000} = 37.493^\circ$, with a positional radius uncertainty of $3\arcmin$ at a 90\% confidence level \citep{Cheng2025GCN}. Subsequent observations with EP-FXT provided an improved localization with an uncertainty of ${\sim}10\arcsec$ \citep{Cheng2025GCN}. 
Following the detection, multi-wavelength follow-up observations were conducted to search for an optical/NIR counterpart \citep{He2025GCN,Lipunov2025GCN,Hall2025GCN,Eyles-Ferris2025GCN,Bochenek2025GCN}. Thanks to the rapid response the ALT100C telescope reported a candidate optical counterpart consistent with the EP-FXT localization \citep{He2025GCN}. 
However, this candidate has not been confirmed by contemporaneous observations \citep{Lipunov2025GCN,Hall2025GCN,Eyles-Ferris2025GCN,Bochenek2025GCN}. Indeed, a reanalysis found that the reported candidate was spurious.

In this paper, we report on our multi-wavelength dataset of the FXT EP250905a. We present extensive imaging and spectroscopic observations spanning from $\sim$1 to 150 days after the EP-WXT trigger. In Sect.~\ref{sec:observations}, we describe the multi-wavelength observations used in this work. In Sect.~\ref{sec:results}, we present our main results. Sect.~\ref{sec:interpre} provides an interpretation of the EP250905a emission, and in Sect.~\ref{sec:conclusion} we summarize and conclude our work. 
Throughout this paper, magnitudes are reported in the AB system, and we adopt a $\Lambda$CDM cosmology with $H_0 = 67.66$~km~s$^{-1}$~Mpc$^{-1}$ and $\Omega_{\Lambda} = 0.69$ \citep{Planck2020}.

\begin{figure}
\centering
\includegraphics[scale=0.76]{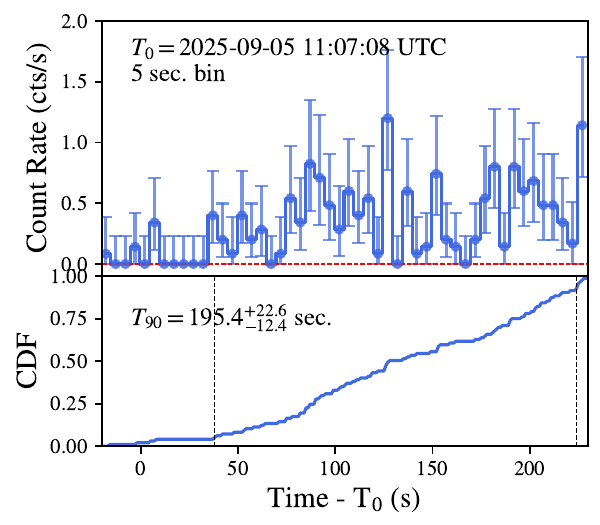}
\vspace{-0.3cm}
\caption{\emph{Top panel:} 0.5--4~keV background-subtracted light curve of EP250905a obtained by the EP-WXT instrument. Time is in seconds, and the bin size is 5~s. The uncertainties are at $1\sigma$ using the \citealt{Kraft1991} statistic. \emph{Bottom panel:} cumulative distribution function (CDF) of the counts detected by EP-WXT. The $T_{90}$ duration of $195.4_{-12.4}^{+22.6}$~sec. is indicated by the vertical dashed lines. The observation was stopped before the end of the transient, so we consider the duration to be a lower limit.}
\label{fig:LC-wxt}
\end{figure}

\section{Observations} \label{sec:observations}

Several multi-wavelength ground- and space-based telescopes observed the transient EP250905a over the period ranging from $\sim0.1$ to $150$~days after the X-ray trigger. All the optical and NIR photometry is calibrated to Panoramic Survey Telescope \& Rapid Response System \citep[Pan-STARRS;][]{Chambers2016,Flewelling2018} or Two Micron All Sky Survey  \citep[2MASS;][]{Skrutskie2006} stars in the field, respectively, and are provided in Table~\ref{tab:photometry}; the X-ray and radio fluxes are in Tables~\ref {tab:x-rays} and \ref{tab:radio}, respectively.

\subsection{Einstein Probe (EP) observations}

\subsubsection{EP -- Wide-field X-ray Telescope (EP-WXT)}

EP250905a was initially detected by the EP-WXT instrument. The X-ray event data were reduced with a dedicated chain tool pipeline \texttt{wxtpipeline} and the corresponding calibration database (CALDB) developed for EP-WXT (Liu et al., in prep.). The CALDB incorporates results from on-ground and in-orbit calibration observations \citep{Cheng2024}. Analysis of the processed EP-WXT data indicates that the flare began at approximately $T_0=$2025-09-05 11:07:08~UTC

The light curve is extracted using the \texttt{XSELECT} tool \citep{NASA2014}, considering a circular region of radius 9\arcmin, while the background was estimated from a source-free annular region centered on the source with inner and outer radii of 11\arcmin\ and 20\arcmin, respectively. Fig.~\ref{fig:LC-wxt} depicts the EP-WXT light curve (background subtracted) in the energy range of 0.5--4~keV (top panel), at a time resolution of 5~sec. Moreover, the $T_{\rm 90}$ duration of the transient, which was computed based on the cumulative distribution function (CDF; Fig.~\ref{fig:LC-wxt}, bottom panel) of the X-ray photons, is $T_{\rm 90}=195_{-12}^{+23}$~seconds. Due to the fact that the observation was stopped before the end of the transient (for pointing the EP-FXT instrument at the transient), the duration is a lower limit of the event. The light curve shows a clear short-term variability.

In order to extract the X-ray spectrum, the ancillary and response files were obtained following standard procedures using the \texttt{XSELECT} tool \citep{NASA2014}, considering the same regions defined for the light curve. 
To obtain the X-ray spectral parameters, we used the spectral fitting program \texttt{XSPEC} v12.14 \citep{Arnaud1996}, considering the background-subtracted spectra and the Cash statistic \citep[W-stat;][]{Cash1979} to account for the low number of counts of the target. 
We fit an absorbed power-law model (\texttt{tbabs*pow} model in \texttt{XSPEC}; $dN/dE\propto E^{-\Gamma_{\rm{WXT}}}$) to the spectrum, which was binned to have at least one photon per bin, where the \texttt{tbabs} model describes the Galactic absorption\footnote{We also test whether an additional intrinsic absorption component is statistically significant. Model comparison using the AIC (Akaike Information Criterion) and BIC (Bayesian Information Criterion) shows no meaningful preference for including intrinsic absorption ($\Delta{\rm AIC} = 1.54$ and $\Delta{\rm BIC} = 0.88$), indicating that the additional component is not statistically required.}. For the extraction of the X-ray spectral parameters, we fixed the Galactic hydrogen column density (atomic and molecular) to $7\times$10$^{20}$~cm$^{-2}$~as determined for this line-of-sight  \citep{Kalberla2005, Kalberla2015, Willingale2013}. 
The best-fit power law index and unabsorbed average flux are $\Gamma_{\rm{WXT}}=1.7\pm0.4$ and $F_X^{\rm{unabs}}$=$\left(4.2\pm1.1\right)\times10^{-10}$~erg~cm$^{-2}$~s$^{-1}$ at 0.5--4~keV (at 90\% confidence level), respectively, for a W-stat of 49.96 for 81 degrees of freedom (dof) at 0.5--4~keV, and a peak flux of $F_{\rm{X,peak}}\approx1\times10^{-9}$~erg~cm$^{-2}$~s$^{-1}$ (and $\sim2\times10^{-9}$~erg~cm$^{-2}$~s$^{-1}$ at 0.3--10~keV). Fig.~\ref{fig:SP-wxt} shows the WXT spectra and the best-fit absorbed power-law model.

\begin{figure}
\centering
\includegraphics[scale=0.4]{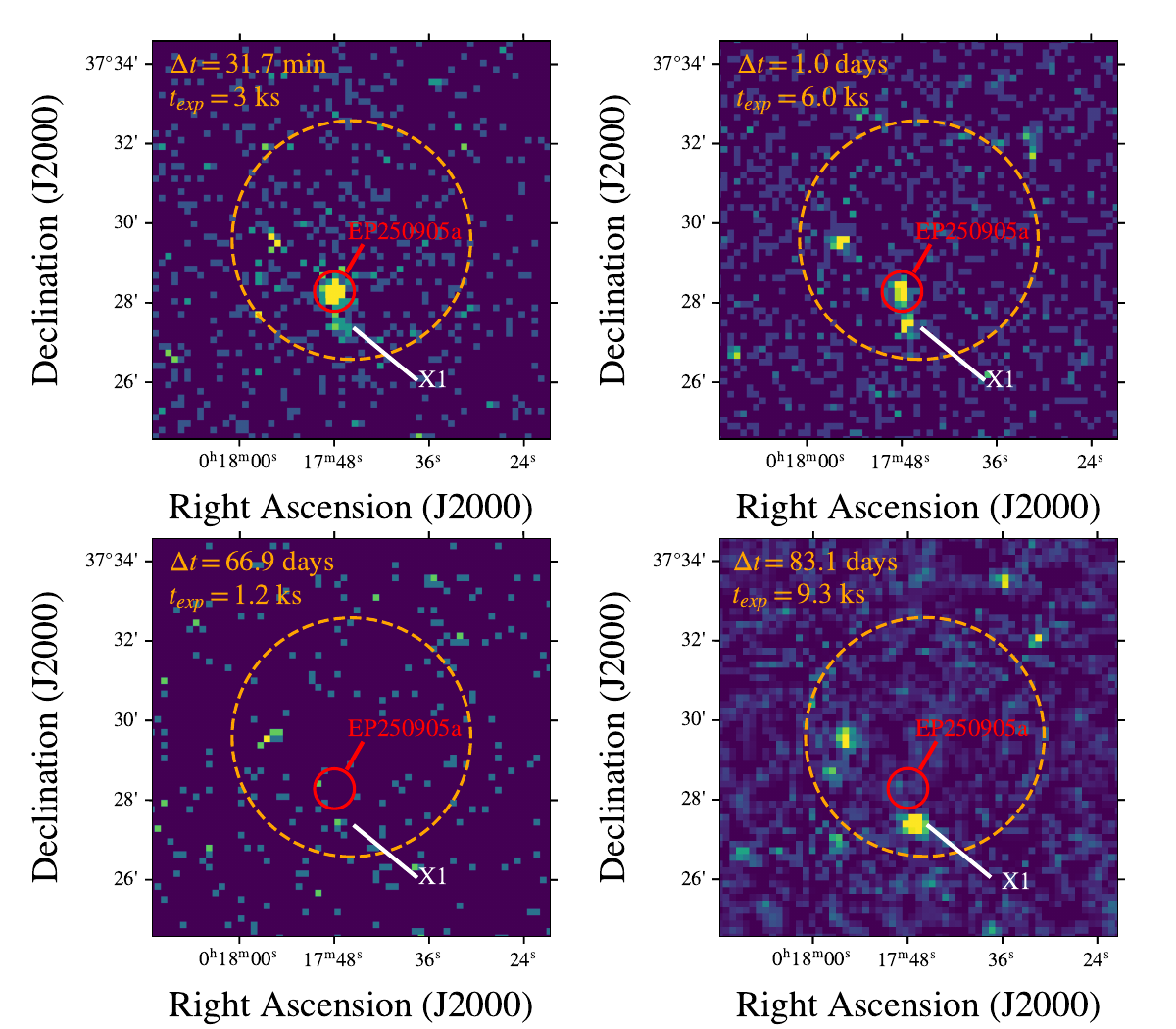}
\vspace{-0.2cm}
\caption{Combined EP-FXTA and FXTB image of EP250905a during four different epochs. The dashed orange circle indicates the EP-WXT positional uncertainty of $3\arcmin$ (90\% confidence level), while the red circle depicts the position of the X-ray afterglow of EP250905a detected by EP-FXT. The persistent X-ray source, labeled as X1, is located near the counterpart of EP250905a (at ${\approx}55$~arcsec).}
\label{fig:EP-FXT}
\end{figure}

\subsubsection{EP-Follow-up X-ray Telescope (EP-FXT)} \label{sec:FXT}

The EP-FXT instrument consists of two modules, systems A and B (each with its own detector), and is sensitive over the 0.5–10 keV energy band \citep{Yuan2015,Yuan2018,Yuan2022}. Four follow-up EP-FXT observations of EP250905a were performed.
The first observation began on 2025 September 5 at 11:13:53 UTC, with an exposure time of 3 ks, corresponding to $\Delta t = 405$~s after the EP-WXT trigger. The results of this initial observation were also reported by \cite{Wang2025GCN250905a}. The second observation started on 2025 September 6 at 10:22:38 UTC, with an exposure time of 6 ks, corresponding to $\Delta t = 0.97$~days after $T_0$.
Two additional epochs were obtained in the following months 
as part of a target-of-opportunity (ToO) program (Quirola-Vásquez PI). The third epoch was carried out on 2025 November 11 at 09:10:26 UTC, with an exposure time of 1.2 ks, corresponding to $\Delta t=67$~days after the trigger. The fourth epoch was performed on 2025 November 27 at 12:09:41 UTC, with an exposure time of 9.3 ks, corresponding to $\Delta t=83$~days after $T_0$.

We reduced the EP-FXT data using the dedicated data analysis pipeline \citep[\texttt{fxtsoftware} v1.20\footnote{Available at http://epfxt.ihep.ac.cn/analysis};][]{Zhao2025}, following the EP-FXT User’s Guide (v1.20). Fig.~\ref{fig:EP-FXT} shows the combined EP-FXT module A and B images of EP250905a for the four epochs.
Because the transient is located in close proximity to a persistent X-ray source labeled as X1 in Fig.~\ref{fig:EP-FXT}, we investigate if the EP-FXT localization of the source originally reported by \cite{Wang2025GCN250905a} is affected by the presence of this source. To this end, we used the \texttt{CIAO} software package (version 4.16; \citealt{Fruscione2006}) and the \texttt{wavdetect} algorithm \citep{Freeman2002} to identify X-ray sources in the EP-FXT data. We then applied a boresight correction by matching the detected X-ray sources to the \emph{Gaia} DR3 catalog \citep{Gaia_DR3_2020}.
The first EP-FXT observation, corresponding to the epoch when EP250905a was brightest, reveals an X-ray counterpart within the EP-WXT localization region at coordinates 
RA$_{\rm J2000}$ = 00$^{\rm h}$17$^{\rm m}$48$^{\rm s}.0$, 
Dec$_{\rm J2000}$ = +37$^\circ$28$\arcmin$17$\arcsec.4$ (see Fig.~\ref{fig:EP-FXT}), with a positional uncertainty of $\approx 8.5\arcsec$ at the 90\% confidence level, considering both the statistical uncertainty and systematic error of 3.4\arcsec. 
The coordinates reported here are therefore offset by $\approx6.58\arcsec$, and with an enhanced smaller positional uncertainty of $\approx1.5\arcsec$, with respect to those reported by \citet{Wang2025GCN250905a}.

The X-ray spectrum and response files were obtained using the \texttt{XSELECT} tool, considering a circular region around the target with a radius of 40\arcsec; meanwhile, the background region was selected in a nearby source-free circular region with a radius of 90\arcsec. The background-subtracted spectra were binned, requiring at least 1~count per spectral bin, and analyzed using \texttt{XSPEC} v12.14 \citep{Arnaud1996} and using the W-stat statistic \citep{Cash1979}. In the same way as EP-WXT data, we fit an absorbed power-law model, fixing a Galactic hydrogen column density to $7\times$10$^{20}$~cm$^{-2}$ \citep{Kalberla2005, Kalberla2015}.

Figure~\ref{fig:SP-fxt} depicts the EP-FXT spectra and the best-fit absorbed power-law model during the first two epochs. Both epochs show clear detections, the spectral shape is consistent with being constant ($\Gamma_{\rm FXT}^{\rm{E1}}=2.4\pm0.3$ and $\Gamma_{\rm FXT}^{\rm{E2}}=2.3\pm0.5$, respectively). 
On the other hand, EP250905a was not detected by EP-FXT during the third and fourth epochs (see Fig.~\ref{fig:EP-FXT}). The estimated 3$\sigma$ upper limits were computed fixing the photon index to $\Gamma=2.3$. X-ray photometry is available in Table~\ref{tab:x-rays}.

\subsection{\emph{Swift}-XRT}

The \textit{Swift} satellite \citep{Gehrels2004} observed the location of EP250905a starting on September 5 at 12:58:37 UTC (i.e., $\sim1.86$~hours after $T_0$) using the X-ray Telescope (XRT), for a total exposure time of $\sim1.2$~ks.
Using the online data products generator provided by the University of Leicester \citep{Evans2007,Evans2009}\footnote{\url{https://www.swift.ac.uk/user\_objects/}},  EP250905a was not detected. We estimate a $3\sigma$ count-rate upper limit of $\approx0.02$~cts~s$^{-1}$. Assuming a Galactic hydrogen column density of $N_{\rm H}=7\times10^{20}$~cm$^{-2}$ and a power-law spectrum with photon index $\Gamma_{\rm X}=2.3$, and using \texttt{PIMMS} v4.15, we derive a corresponding flux upper limit of $\lesssim7.9\times10^{-13}$~erg~cm$^{-2}$~s$^{-1}$ in the 0.5--10~keV band.

\begin{figure*}
\centering
\includegraphics[scale=0.75]{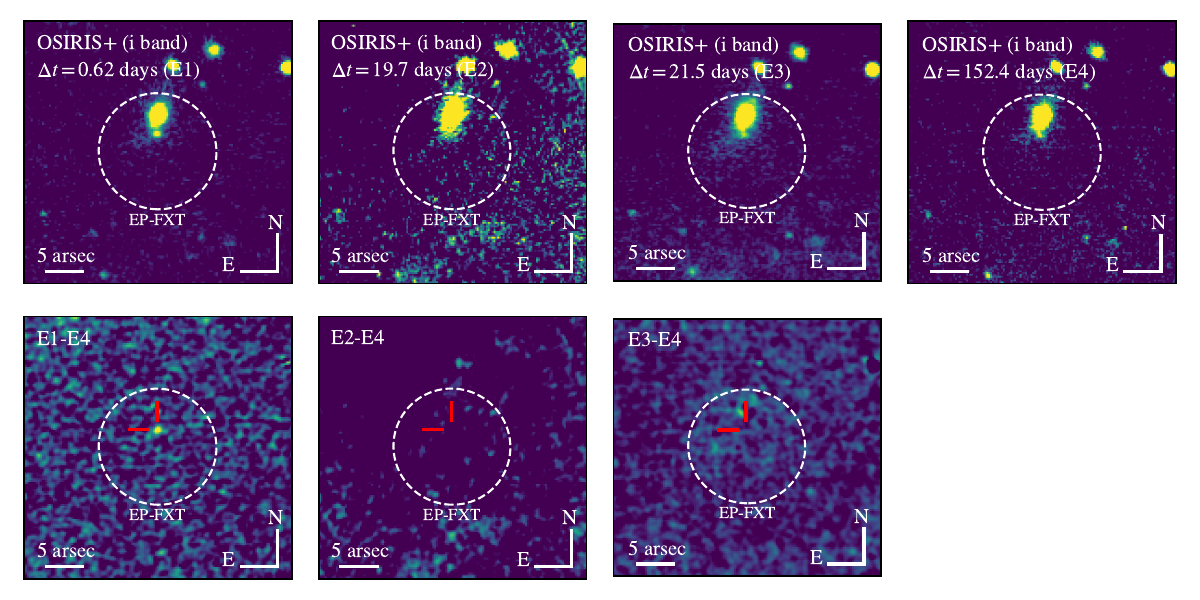}
\vspace{-0.3cm}
\caption{\emph{Top panels:} From left to right, we show our four GTC/OSIRIS+ $i$-band observations centered on the EP-FXT position (dashed circle). The time since $T_0$ in days is indicated in the panels. \emph{Bottom panels:} Difference images derived using the HOTPANTS software suite, where we subtracted the last epoch (rightmost panel of the above row), also known as the template image, from the first three observations (E4). In the E1-E4 difference image, an optical counterpart marked by red tick marks is clearly detected. This counterpart is not detected in subsequent epochs.
}
\label{fig:optical}
\end{figure*}

\begin{figure*}
\centering
\includegraphics[scale=1.0]{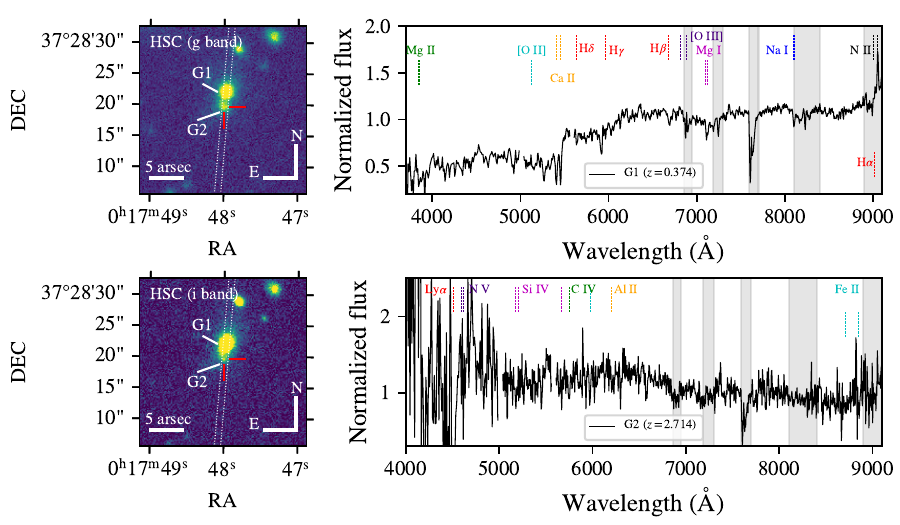}
\vspace{-0.3cm}
\caption{\emph{Left panels:} $g$-band (top) and $i$-band (bottom) HSC images \citep{Tanaka2021} of the field of EP250905a. The two host galaxy candidates are labeled as G1 and G2; the orientation of the 1\arcsec~wide GTC/OSIRIS+ slit is marked with a white-dotted rectangle. The position of the optical counterpart (see Sect.~\ref{sec:opt-counter}) is denoted by red tick marks. \emph{Right panels:} GTC-OSIRIS+ R1000R/R1000B optical spectrum of host galaxy candidates G1 (top panel; ${\approx}19.7$~days after $T_0$) and G2 (bottom panel; ${\approx2}0.7-25.6$~days after $T_0$). For the G1 galaxy, we indicate the location of the H$\alpha$, [O II] and [O III], H$\beta$, H$\gamma$ and H$\delta$ lines consistent with a redshift of $z_{\rm{G1}}=0.374$. For the G2 galaxy (bottom panel), we identify the interstellar medium (ISM) absorption lines Si IV, C IV, Al II, and Fe II, all consistent with a redshift of $z_{\rm{G2}}=2.714$. We also show wavelength regions affected by telluric lines (gray shaded regions). The data was rebinned using \texttt{SpectRes} \citep{Carnall2017} to a resolution of 10~\AA~(R1000R) and 15~\AA~(R1000B) for visibility purposes.}
\label{fig:host_spec}
\end{figure*}

\begin{figure}
\centering
\includegraphics[scale=0.8]{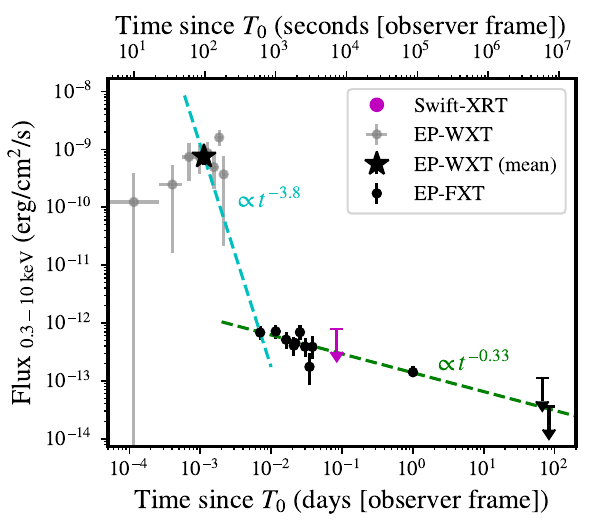}
\vspace{-0.3cm}
\caption{The 0.3-10~keV X-ray light curve of EP250905a. The first data
point is the EP-WXT mean flux (and time), and the next data points are from the EP-FXT (detections and upper limits) and \emph{Swift}-XRT (upper limit) observations. The cyan and green dashed lines indicate the best-fit power laws.}
\label{fig:x_flux}
\end{figure}

\subsection{GTC/OSIRIS+ observations}
\subsubsection{Imaging}

We obtained four imaging epochs of the field of EP250905a using the upgraded Optical System for Imaging and low-intermediate-Resolution Integrated Spectroscopy (OSIRIS+) instrument \citep{Cepa2000} mounted on the Gran Telescopio Canarias (GTC). Our observations were taken using the $i$-band filter. The first observation (E1) of the field started on September 6 at 02:01:39 UTC (i.e., $\Delta t\approx0.62$~days, after the X-ray trigger, and with a seeing of $\sim0.81$\arcsec~FWHM) using an exposure time of $9\times60$~sec.
The second (E2) and third (E3) observations were taken on September 25 and 26 at 03:40:03 and 22:13:34 UTC, respectively (i.e., $\Delta t\approx$19.7 and 21.5~days after $T_0$, and with a seeing of $\sim1.1$ and 0.84\arcsec~FWHM, respectively), and with an exposure time of $9\times60$ and $19\times150$~sec, respectively.
Finally, a last epoch (E4) was taken on February 4 at 20:35:03 UTC (i.e., $\sim152.4$~days after $T_0$, and with an optical seeing of $\sim0.78$\arcsec~FWHM) with an exposure time of $9\times60$~sec.
The data were bias subtracted and flat-field corrected, adopting standard \texttt{PyRAF} tasks \citep{Pyraf2012}, and cosmic rays were removed using the \texttt{LACosmic} task \citep{van_Dokkum2001}. The world coordinate system information is implemented using \texttt{astrometry.net} codes \citep{Lang2010}. Fig.~\ref{fig:optical}, top panels, depict the individual OSIRIS+ $i$-band images centered on the EP-FXT position of EP250905a.

\subsubsection{Spectroscopy}

We obtained three spectroscopic observations of the host galaxy candidates of the transient EP250905a (labeled as G1 and G2; see Sect.~\ref{sec:redshift}), using the OSIRIS+ instrument \citep{Cepa2000}.
The first epoch, designed to get the spectrum of G1, was obtained on September 25 at 04:18:56 UTC (i.e., $\Delta t\approx19.72$~days after $T_0$) using an exposure time of 3$\times$1200~sec, grism R1000R (covering $\approx$5000-9500~\AA), and a 1\arcsec~wide slit placed at the parallactic angle.
The second observation was obtained on September 26 at 02:42:46 UTC (i.e., $\Delta t\approx20.65$ after the X-ray trigger), using an exposure time of 3$\times$1200~sec, grism R1000R, and a 1\arcsec~wide slit oriented to cover G1 and G2 simultaneously.
Finally, a last GTC epoch was taken on October 1 at 01:44:18 UTC (i.e., $\Delta t\approx25.61$~days after $T_0$), with a setting similar to the second epoch, using the grism R1000B (covering $\approx$4000-8000~\AA) instead of the R1000R grism.

\subsection{Nordic Optical Telescope (NOT)}

We observed the target using the Andalucía Faint Object Spectrograph and Camera (ALFOSC) and the Nordic Optical Telescope near-infrared camera and spectrograph (NOTCam), both mounted on the NOT at the Roque de los Muchachos Observatory (La Palma, Canary Islands, Spain). 
We observed the transient position using
NOTCam at five epochs between September 5 and September 8 (i.e., $\Delta t\approx$0.47, 0.52, 0.75, 1.73, and 2.59~days after $T_0$) using the $J$-band filter (program NOT 70-301 and 71-812, PI: Jonker and Fynbo/Xu, respectively).
Meanwhile, ALFOSC observed the field using the $g$ and $r$ bands on October 1 at 01:54:46.8 and 02:11:20.5 UTC (i.e., $\Delta t\approx25.6$~days after the trigger; program NOT 71-805, PI: Jonker).
All observations were carried out using a standard dithering pattern. The data were reduced following standard procedures using the {\sc iraf} package \citep{Tody1986}.

\subsection{Las Cumbres Observatory (LCO)}

EP250905a was observed with a fully robotic 1-m telescope at the Teide Observatory located in the Canary Islands-Spain on September 5 at 22:34:45.9 UTC (i.e., $\Delta t\approx 0.49$~days after $T_0$) using the $r$ filter. The data were reduced using the \texttt{BANZAI} pipeline \citep{McCully2018} and standard procedures with IRAF \citep{Tody1986}.

\subsection{The Thai Robotic Telescope (TRT)}

The TRT is an automated telescope network comprising four 70 cm CDK700 Telescopes equipped with Andor CCD cameras, distributed in the United States (SRO), Chile (CTO), Australia, and China (GAO). EP250905a was observed on September 5 at 11:56:27.3 UTC (i.e., $\Delta t\approx0.04$~days after $T_0$). Images were reduced using standard procedures with IRAF \citep{Tody1986}.

\subsection{The Liverpool telescope}

The IO:O Imager at the robotic 2m Liverpool Telescope \citep{Steele2004},  located at the Observatorio del Roque de los Muchachos, La Palma, was triggered for follow-up of EP250905a four times. The first three observations were obtained using the $gri$ filters on September 5 ($\Delta t\approx0.47$~days after $T_0$), and were reported by \citet{Bochenek2025GCN, Eyles-Ferris2025GCN}. A fourth observation was obtained $\Delta t\approx25.6$~days after $T_0$ using the 
$i$ filter.

\subsection{ALT100C telescope}

The transient was observed using the Altay 1-Meter Telescope (ALT-100C), part of the JinShan project, located at Altay, Xinjiang, China. The transient was observed seven times using the $r$-, $i$-, and $z$-band filters at $\Delta t\approx$0.15, from $\approx$0.43 to 42.3, and $\approx$0.28~days after $T_0$, respectively. The ALT-100C images were reduced using standard procedures with IRAF \citep{Tody1986}, including bias subtraction, flat-fielding, and image combination.

\subsection{VLT/XSHOOTER}

X-shooter \citep{Vernet2011} observed the host galaxy candidate G2 of FXT EP250905a (see for more details Sect.~\ref{sec:redshift}) on September 30 at 03:52:47 UTC (i.e., $\Delta t\approx24.7$~days since the X-ray trigger). The instrument was operated in nodding mode using an ABBA sequence to improve the sky subtraction (ESO 114.27PZ.010, PI Tanvir), with $2\times1440$~s exposures in the UVB and VIS arm and $6\times480$~s exposures in the NIR arm. Seeing was $\sim0.39\arcsec$~during these observations.
The data was reduced with the ESO X-shooter pipeline \citep{Goldoni2006,Modigliani2010}.

\subsection{MeerKAT observations}

The field of EP250905a was observed with the MeerKAT radio telescope over four epochs at $\Delta t \approx 21.4$, 40.4, 61.4, and 104.2 days after $T_0$, using a central frequency and bandwidth of 3 GHz and  875~MHz, respectively, and 44 minutes of on-source time per epoch. All of our MeerKAT observations were reduced using the \texttt{oxkat} pipeline \citep{oxkat}, a set of semi-automated scripts for the calibration and imaging of MeerKAT data. \texttt{oxkat} performs phase reference calibration using \texttt{CASA} \citep{casa1,casa2}, self calibration using \texttt{CubiCal} \citep{cubical}, and imaging using \texttt{WSClean} \citep{wsclean}.
We did not detect a radio counterpart in any of the epochs, with a $3\sigma$ upper limit of $<30$~$\mu$Jy/beam in all cases (see Table~\ref{tab:radio}).

\subsection{Fermi-GBM monitoring}\label{sec:gamma}

At the time of the EP trigger, the Gamma-Ray Burst Monitor (GBM) on board the \emph{Fermi} satellite had full temporal and spatial coverage of the localization region, but there was no corresponding onboard trigger.
We performed a \emph{Fermi}-GBM targeted search in the time interval [$T_0-50$;$T_0+500$] s using the method reported in \citet{Goldstein2019}, finding no corresponding signal. 
Assuming a ''soft'' Band function \citep[with $E_{\rm peak}=70$~keV, $\alpha=-1.9$, and $\beta=-3.7$;][]{Band1993}, we derived a $3\sigma$ flux upper limit of $<7.06\times10^{-9}$~erg~s$^{-1}$~cm$^{-2}$ in the energy range 10--1000~keV over the full duration of the burst.
Fig.~\ref{fig:high_sed} depicts the high-energy spectral energy distribution of EP250905a covering the X-ray and gamma-ray range by the EP-WXT detections and the upper limit from \emph{Fermi}-GBM.

\section{Results} \label{sec:results}

\subsection{Optical and NIR counterpart} \label{sec:opt-counter}

We performed image subtraction using the High Order Transform of PSF and Template Subtraction code \citep[HOTPANTS;][]{Becker2015}, adopting the OSIRIS+ E4 epoch as the host template. We effectively, therefore, assume that no emission from the transient was detected in that observation. The E4 image is subtracted from the images of the E1, E2, and E3 epochs. The resulting subtractions are shown in the bottom panels of Fig.~\ref{fig:optical}. 
The subtraction E1$-$E4 reveals a significant residual within the EP-FXT error region, with a signal-to-noise ratio of $\sim$\,5.9. The source is located at
RA$_{\rm J2000}^{\rm opt} = 00^{\rm h}17^{\rm m}$48\fs00,
Dec$_{\rm J2000}^{\rm opt} = +37^\circ28\arcmin$19\farcs6,
with an astrometric uncertainty of $\approx$\,0\farcs1, estimated using \texttt{SExtractor} \citep{Bertin1996}.
The transient has an AB magnitude of $m_i = 23.23 \pm 0.18$, calibrated against Pan-STARRS. 

We independently confirm this detection using the \texttt{PyZOGY} package \citep{Zackay2016}, adopting an archival $i$-band image from the Canada-France-Hawaii Telescope (CFHT) as the reference template. 
Given the spatial and temporal coincidence between this optical transient source and the X-ray detections of EP250905a, we identify this target as the optical counterpart of EP250905a.
No significant residual is detected in the E2$-$E4 and E3$-$E4  images (see Fig.~\ref{fig:optical}, bottom panels). We derive $3\sigma$ upper limits of $m_i > 23.5$ and $m_i > 25.3$~AB~mag for these epochs, respectively, confirming the fading of the counterpart discovered in the E1$-$E4 image.

We applied the same procedure and template to the $i$-band observations obtained with the ALT100C telescope. However, no counterpart was detected in these observations.
In the NIR, we followed a similar procedure to search for a counterpart in the $J$-band images obtained with the Nordic Optical Telescope, using HOTPANTS for image subtraction. The final epoch, taken with NOTCam (i.e., $\approx$\,2.6~days after $T_0$), was used as the template. No counterpart was detected in the residual images.

Spectroscopic observations obtained $\sim$\,20.7~days after $T_0$ cover the position of the optical counterpart (see Fig.~\ref{fig:host_spec}, left panels), but no spectral features associated with the transient are detected (see Sect.~\ref{sec:redshift}), consistent with the E2$-$E4 non-detection.

\subsection{Host galaxy and redshift estimation} \label{sec:redshift}

We have no direct redshift measurement of the optical counterpart. However, because of the localization of the optical counterpart and the EP-FXT position, there are two galaxies that we identify as the candidate host galaxy of EP250905a. The left panels of Figure~\ref{fig:host_spec} show $g$- and $i$-band Hyper Suprime-Cam (HSC) Legacy Archive images \citep{Tanaka2021} of the field of EP250905a, where the two host candidates are labeled as G1 and G2. Using the archive HSC images and \texttt{SExtractor} \citep{Bertin1996}, we measure the G1 and G2 coordinates as:
RA$_{\rm J2000}^{\rm{G1}}$ = 00$^{\rm h}$17$^{\rm m}$47\fs97,
Dec$_{\rm J2000}^{\rm{G1}}$ = +37$^\circ$28$\arcmin$22\farcs14, 
and 
RA$_{\rm J2000}^{\rm{G2}}$ = 00$^{\rm h}$17$^{\rm m}$48\fs00,
Dec$_{\rm J2000}^{\rm{G2}}$ = +37$^\circ$28$\arcmin$19\farcs87.

For each galaxy, we estimate the probability that it falls close to the position of the optical counterpart of EP250905a by chance using the formalism of \citet{Bloom2002}\footnote{The chance alignment probability $P_{\rm ch}$ quantifies the likelihood that a galaxy of a given magnitude or brighter is located within the observed angular separation from the transient by chance given the projected surface density of galaxies as a function of their magnitude.}. 
Using the Pan-STARRS magnitude for G1 ($m_r = 20.39 \pm 0.05$~AB~mag) and the derived magnitude for G2 ($m_i = 22.15 \pm 0.06$~mag using a circular aperture with a $1\arcsec$ radius, and assuming a color in the range of $m_r - m_i \approx 0.0$--$1.0$~AB~mag) and offsets to G1 and G2 of \,2\farcs57$\pm$0\farcs38 and 0\farcs27$\pm$0\farcs38, respectively, we derive chance alignment probabilities of $P_{\rm ch} = 6 \times 10^{-3}$ and $P_{\rm ch} \approx (3$--$6) \times 10^{-4}$ for G1 and G2, respectively.

We determine the redshift of the two host galaxy candidates from late-time spectroscopic observations.
Figure~\ref{fig:host_spec}, right panels, depicts the GTC/OSIRIS+ R1000R and R1000B grating combined and normalized spectra (regarding their mean value) of G1 (top) and G2 (bottom). In the case of G1 (see Fig.~\ref{fig:host_spec}, right-top panel), we detect weak evidence for the presence of emission lines such as 
H$\alpha$ and N II$\lambda6584$,
and for the strong absorption lines Mg~I$\lambda5167,5184$, Mg~II$\lambda2796,2803$, Ca~II$\lambda3935,3970$, Na~I$\lambda5892,5898$~\AA, H$\delta$, H$\gamma$, and H$\beta$ all at a common redshift of $z_{\rm{G1}}=0.374$.
On the other hand, the host candidate G2 (see Fig.~\ref{fig:host_spec}, right-bottom panel) shows some absorption lines such as Ly$\alpha$, N~V$\lambda1238,1242$, Si~IV$\lambda1393,1402,1527$, C~IV$\lambda1548,1550$, Al~II$\lambda1670$, and Fe~II$\lambda1608,2382,2344$~\AA~all at a common redshift of $z_{\rm{G2}}=2.714$. These lines were also observed in the VLT/XSHOOTER spectra (see Fig.~\ref{fig:xshooter}), along with the tentative detection of one emission line at $\sim$18,610~\AA\ which is consistent with [O III]$\lambda5007$\AA~(the line does not show any signs of a broad component).
It was the emission line from Xshooter that made it possible to confirm the redshift, together with Ly$\alpha$.

\subsection{Optical/NIR light curves}

Based on the redshifts of the candidate host galaxies, we derive the absolute optical magnitude of the counterpart. Using observations obtained at $\approx 0.62$ days after $T_0$, we find $M_i \approx -17.8$ and $M_i \approx -22$ AB mag, assuming association with G1 and G2, respectively. No additional detections are found in earlier or later optical and NIR observations; however, we place deep upper limits, particularly in the $i$-band (see Table~\ref{tab:photometry}).

\subsection{X-ray light curve}

The EP-WXT light curve (see Fig.\ref{fig:LC-wxt}) has a duration of $\approx200$~sec, and does not show a single-pulse profile, but rather a multi-peaked, slowly evolving one.
In Figure~\ref{fig:x_flux} we show the 0.3--10~keV X-ray light curve of EP250905a in the observer frame. The first data point (black star) corresponds to the mean EP-WXT flux (at the mid time of $\approx100$~s), followed by the EP-FXT epoch 1 detection and \emph{Swift}-XRT upper limits. 
There appears to be a steep ($\sim$3 orders of magnitude) decay in X-ray flux between both the individually binned and mean EP-WXT flux points and the flux at the start of the first EP-FXT observation. This drop can be described by a power law with index $\approx-3.8$, which we consider as an upper limit, since it could be even steeper. Considering only the EP-FXT epoch 1 and epoch 2 observations ($\delta t$$\approx$\,$10^{-2}$--$10^{0}$ days), the power law decay index is significantly less steep, with an index of $=-0.33\pm0.06$, indicating a sharp break in the temporal behavior. The two power-laws are shown in Fig.~\ref{fig:x_flux} by the cyan and green dashed lines.  The 3$\sigma$ upper limits for epochs 3 and 4 remain marginally consistent with an extrapolation of this power-law slope, although the strength of the epoch 4 limit hints at a possible late-time steepening decline.

\begin{figure}
\centering
\includegraphics[scale=0.52]{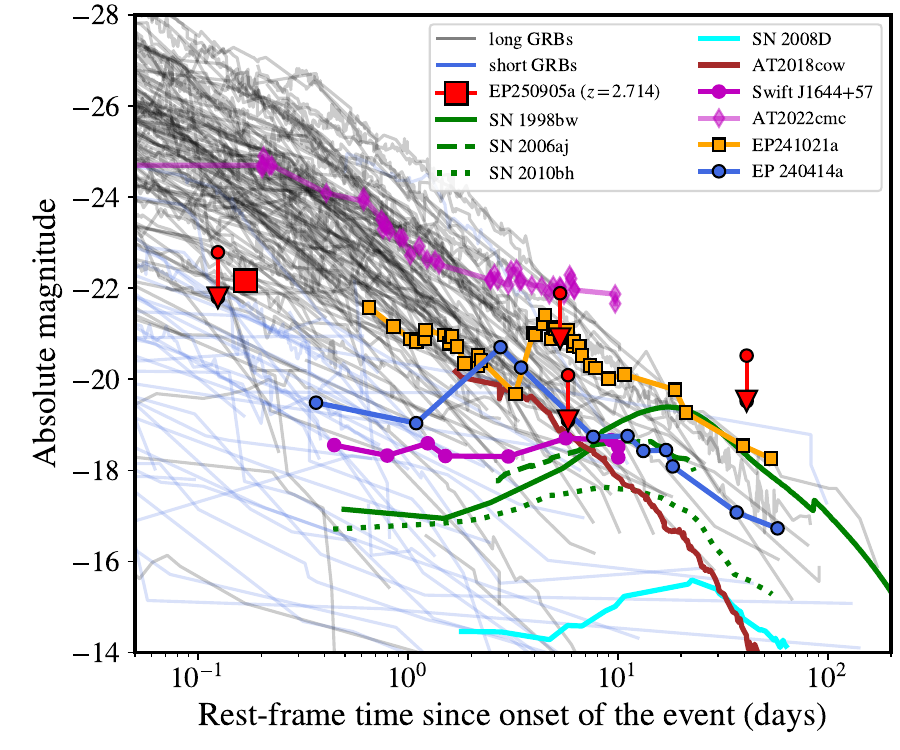}
\vspace{-0.3cm}
\caption{
$i$-band light curve comparison of EP250905a and a sample of transients, such as long and short GRBs \citep{Kann2006,Kann2010,Kann2011,Nicuesa2012,Kann2024}, the SN SBO event XRF~080109/SN~2008D, the X-ray flashes XRF~100316D/SN~2010bh and SN~2006aj/XRF~060218, the broad-line Ic SN~1998bw \citep{Clocchiatti2011}, the LFBOT AT2018cow \citep{Xiang2021}, the jetted TDEs AT2022cmc and \emph{Swift}~J1644+57 \citep{Levan2011,Andreoni2022}, and the EP-collapsar FXTs EP240414a \citep{van_Dalen2024} and EP241021a \citep{Quirola2026}. The squares and triangles depict detections and $3\sigma$ upper limits, respectively.
}
\label{fig:optical_comparison}
\end{figure}

\begin{figure}
    \centering
    \includegraphics[scale=0.55]{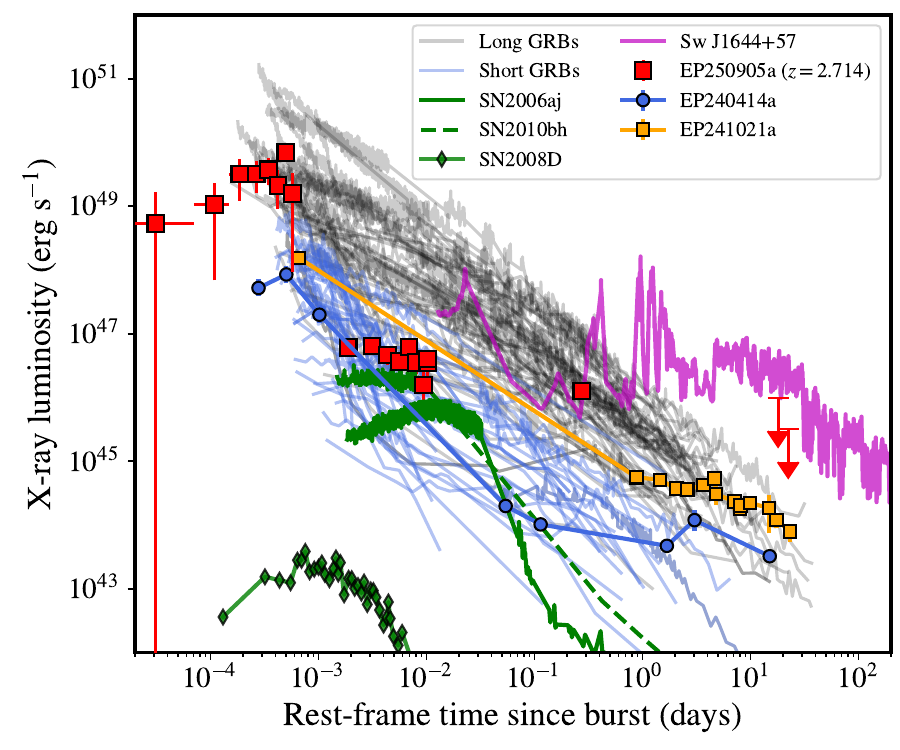}
    \vspace{-0.3cm}
    \caption{
    X-ray light curve of EP250905a (0.3--10 keV) compared with those of the X-ray afterglow light curves of long and short GRBs \citep{Evans2007,Evans2009,Bernardini2012,Lu2015}, the relativistically beamed TDE \emph{Swift}~J1644$+$57 \citep{Bloom2011,Levan2011}, the X-ray flashes XRF~060218/SN~2006aj, XRF~100316D/SN~2010bh \citep{Barniol2015,Starling2011,Modjaz2009,Evans2009,Evans2007,Campana2006}, and XRF~080109/SN~2008D \citep{Soderberg2008}, and the FXTs EP240414a \citep{Sun2025,van_Dalen2024} and EP241021a \citep{Shu2025}. The squares and arrows depict detections and $3\sigma$ upper limits, respectively.
    }
    \label{fig:x_ray_comparison}
\end{figure}

\begin{figure}
    \centering
    \includegraphics[scale=0.55]{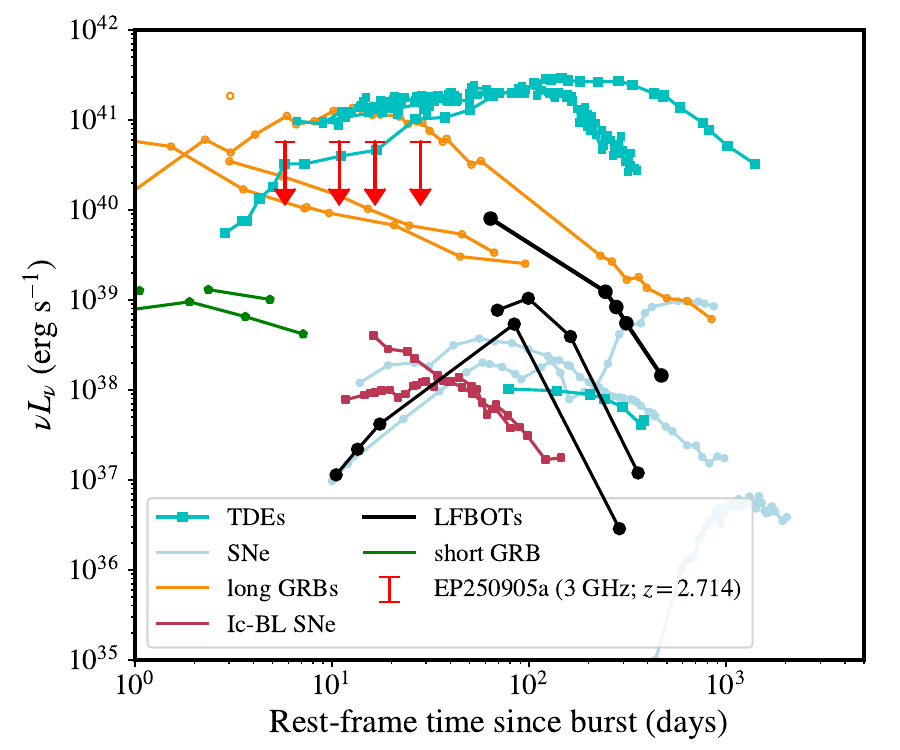}
    \vspace{-0.3cm}
    \caption{
    Radio light curve comparison of EP250905a and the low-frequency (1-10 GHz) light curves of different classes of energetic explosions such as TDEs \citep{Zauderer2011,Berger2012,Zauderer2013,Alexander2016,Eftekhari2018}, SNe exploding in dense circumstellar medium \citep{Soderberg2005,Salas2013} and normal SNe \citep{Weiler1986}, relativistic Ic-BL SNe \citep{Kulkarni1998,Soderberg2010}, long GRBs \citep{Berger2003,Hancock2012,Perley2014,van_der_Horst2014}, and LFBOTs \citep{Margutti2019,Coppejans2020,Ho2020b}. The triangles depict $3\sigma$ upper limits.
    }
    \label{fig:radio_comparison}
\end{figure}

\begin{figure}
    \centering
    \includegraphics[scale=0.7]{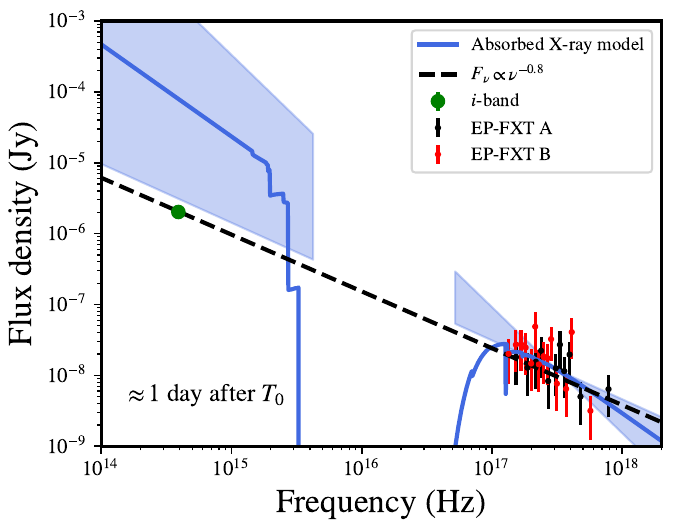}
    \vspace{-0.3cm}
    \caption{The optical to X-ray spectral energy distribution of EP250905a taken at $\sim$\,1~day after $T_0$. The solid blue line represents the best-fit absorbed X-ray power-law model (see Sect.~\ref{sec:FXT}) extrapolated to optical wavelengths. We fit a power-law index of $\beta_{\rm{OX}}=-0.80\pm0.03$ between the optical and the X-ray data (dashed line). The early $i$-band OSIRIS+ detection (the error is smaller than the marker) is marginally consistent with an extrapolation of the unabsorbed X-ray spectrum represented by the blue shaded region at 90\% confidence level. }
    \label{fig:sed}
\end{figure}

\subsection{Spectral Energy Distribution (SED)}

We investigate the optical-to-X-ray SED of EP250905a during our optical detection and the EP-FXT detection (modules A and B) closest in time (the time between both observations is $\Delta t\approx8.9$~hours). First, we correct the photometry for line-of-sight Galactic extinction, adopting $E(B-V)=0.052$~mag \citep{Schlafly2011} and the extinction law of \citet{Calzetti2000}. For the $i$-band filter, this corresponds to a Galactic extinction of $A_\lambda=0.092$~mag. 
Assuming a single power law to the optical-to-X-ray data, we compute a broadband spectral index of $\beta_{\rm OX}=-0.80\pm0.03$, where $F_\nu \propto \nu^{\beta}$.

\section{Discussion and interpretation}\label{sec:interpre}

\subsection{Host galaxy association and redshift assumption}

Despite being the fainter of the two galaxies, the closer proximity of G2 to the optical counterpart position makes it less likely than G1 to be a chance alignment. Moreover, the derived chance alignment probabilities (see Sect.~\ref{sec:redshift}), in which the $P_{\rm ch}$ of G2 is approximately one order of magnitude lower than that of G1, further support G2 as the most likely host galaxy of the transient EP250905a. Therefore, throughout this work, we adopt the transient redshift as $z_{\rm G2}=2.714$.

\subsection{Counterpart interpretation}

We first discuss the energetics of the event, taking into account the upper limit in the $\gamma$-ray band we derived in Sect.~\ref{sec:gamma}.
The $k$-corrected isotropic-equivalent peak luminosity and total energy are $\lesssim5.4\times10^{48}$~erg~s$^{-1}$ and $\lesssim8.9\times10^{50}$~erg, respectively.
Comparing both parameters with a sample of GRBs \citep[e.g.,][]{Tsvetkova2017}, the isotropic-equivalent peak luminosity is consistent with the GRB sample, although it lies in the low-luminosity tail of the distribution.
The high-energy SED of the transient (see Fig.~\ref{fig:high_sed}) depicts the \emph{Fermi}-GBM upper limits, which are not very constraining. Indeed, they lie $\sim$\,1\,dex above an extrapolation of the best-fit photon index $\Gamma_{\rm WXT}=1.7\pm0.4$ to the EP-WXT spectrum, and hence do not permit a clear identification of any peak energy between $\sim$5--1000\,keV, i.e., the source might in principle have been significantly brighter in hard X-rays/gamma-rays than in soft X-rays. This leaves considerable uncertainty in EP250905a's location on the so-called Amati relation. 

In the optical regime, Fig.~\ref{fig:optical_comparison} shows the $i$-band optical light curve of EP250905a compared with several examples drawn from the GRB, SN, jetted TDE, FXT, and LFBOT transient classes.
EP250905a occupies a region populated by multiple classes of transients. At this redshift, the late-time upper limits do not rule out the presence of an SN bump, for instance.  In a GRB context, at $z_{\rm G2}=2.714$, EP250905a falls within the lower luminosity range of long GRBs and remains compatible with their late-time evolution (see Fig.~\ref{fig:optical_comparison}).
 
The corresponding early EP-WXT peak luminosity is $8.7\times10^{49}$~erg~s$^{-1}$. 
After the early EP-WXT prompt emission, the X-ray evolution resembles the canonical GRB X-ray afterglow emission (see Fig.~\ref{fig:x_flux}). In the context of GRB X-ray afterglows, an early steep decay phase is commonly observed \citep{Tagliaferri2005}. This temporal decay slope is steep, typically with a slope in the range of ${\sim}-3$ to ${\sim}-10$, which can be interpreted as the tail or high-latitude emission of the prompt emission \citep{Barthelmy2005b}. In GRBs, following the early steep decay phase, a break occurs around \hbox{$\sim$\,$10^2$--$10^3$~seconds}, where canonical GRBs typically transit to a shallower decay phase, with a power law slope index in the range of $\sim0$ to $\sim-0.7$. Such behavior is commonly interpreted within the standard external forward-shock model \citep{Tagliaferri2005,Nousek2006,Zhang2006,Panaitescu2006} by invoking continuous energy injection into the blast wave, either from a long-lived central engine \citep[an interpretation also proposed for EP241026b; Van Dalen et al., in prep.;][]{Dai1998,Rees1998,Zhang2001,Metzger2011,Bucciantini2012,Rowlinson2013}, or by the stratification of the ejecta Lorentz factor in an impulsively ejected fireball \citep[e.g.,][]{Sari2000,Kumar2000,Nousek2006,Granot2006b,Uhm2012}. Alternative explanations include geometrical effects arising from an off-axis viewing angle \citep[e.g.,][]{Granot2002,Eichler2006,Beniamini2020} and the evolution of shock microphysical parameters with time \citep[e.g.,][]{Fan2006,Granot2006,Ioka2006}.
Finally, the latest X-ray upper limit lies marginally above the extrapolated shallow-decay trend, potentially indicating an acceleration of the rate of decay before this epoch, as commonly observed in GRB afterglows when the plateau phase transitions into a normal decay phase \citep[with a slope of $\sim-1$;][]{Zhang2006}.

Figure~\ref{fig:x_ray_comparison} compares the X-ray light curve of EP250905a with a variety of other X-ray transients.
The peak luminosity disagrees with SN SBOs, such as SN~2008D, which is an order of magnitude fainter.
Its peak luminosity is inconsistent with the majority of X-ray flashes and previously reported FXTs \citep[e.g.,][]{Quirola2022,Quirola2023}. In particular, while EP250905a approaches the luminosity of \emph{Swift}~J1644+57 at late times, it is fainter during the early phase. Finally, it is important to mention that both the peak luminosity and the overall temporal evolution are consistent with long GRBs, in agreement with the optical constraints at the same assumed redshift.

At radio frequencies, Fig.~\ref{fig:radio_comparison} shows the rest-frame radio luminosity evolution ($\nu L_\nu$) for EP250905a (at 3~GHz) compared with several classes of transients, including long and short GRBs, TDEs, CC-SNe exploding in a dense circumstellar medium, Ic-BL CC-SNe, and LFBOTs. 
EP250905a's radio upper limits correspond to $\nu L_\nu \lesssim 7 \times 10^{40}$~erg~s$^{-1}$. These limits do not exclude the typical luminosity range of long GRBs or some relativistic TDEs. They only rule out the most extreme radio-bright transients, such as the relativistic TDEs like AT2022cmc.

Figure~\ref{fig:sed} depicts the SED at $\approx1$~day after $T_0$, together with the extrapolation of the X-ray spectrum to optical frequencies. 
Comparing the single optical detection with the X-ray extrapolation, the optical flux density appears slightly fainter, but remains broadly consistent within the uncertainties, suggesting the lack of a spectral break between optical and X-ray frequencies. 
The optical-to-X-ray SED at $\approx1$~day is well described by a single power law with $\beta_{\rm OX}\approx-0.8$ (see Fig.~\ref{fig:sed}, dashed line). 
Given this $\beta_{\rm OX}$ the source is not classified as an optically dark burst \citep[defined as $\beta_{\rm OX}<0.5$ under $F_\nu\propto\nu^{-\beta}$;][]{Jakobsson2004}. This value is consistent with those typically observed in GRB afterglows \citep[e.g.,][]{Melandri2008}, and likely implies a non-thermal nature of the emission. The measured $\beta_{\rm OX}$ during the early epoch can be compared with predictions of canonical forward-shock (FS) models \citep[e.g.,][]{Sari1999b,Granot2002b,Zhang2006,Gao2013b}. 
Here, we assume that at $\approx1$~day after $T_0$ the transient is in the decelerating relativistic phase under the slow-cooling regime \citep[see for instance][]{Gao2013b}.
In the spectral regime $\nu_m < \nu < \nu_c$, where $\nu_m$ and $\nu_c$ are the characteristic synchrotron and cooling break frequencies, respectively, the spectral index follows $\beta = (1 - p)/2$ \citep[where $p$ is the electron energy distribution index;][]{Gao2013b}. Using $\beta =\beta_{\rm{OX}}\approx -0.80$, we infer $p \approx 2.6$, which lies within the typical range observed in GRB afterglows \citep[e.g.,][]{Panaitescu2002,Curran2010}. This supports a synchrotron origin consistent with a relativistic forward shock.

We note that the interpretation depends critically on the assumed redshift; therefore, we also briefly discuss the scenario in which the transient is located at $z_{\rm G1}=0.374$.
At $z_{\rm G1}$, the early optical detection ($M_i \approx -17.8$~AB~mag), together with the late-time upper limits, excludes the presence of an SN component, as well as jetted TDEs and luminous LFBOTs.
Furthermore, the radio upper limits rule out the presence of a powerful relativistic outflow, while the $\gamma$-ray upper limits are inconsistent with typical GRB emission. 
Overall, under the $z_{\rm G1}$ redshift, the multi-wavelength properties of EP250905a are not consistent with the majority of known transient classes. In this scenario, the low-redshift solution would require an unusually faint optical, X-ray, and radio counterpart. This disfavors the association of EP250905a with galaxy G1. This conclusion is further supported by the $\sim$\,1~dex lower chance coincidence probability of G2 relative to G1. (see Sect.~\ref{sec:redshift}).
Thus, although the redshift ambiguity prevents a definitive classification, the multi-wavelength evidence strongly favors interpreting EP250905a as an afterglow emission, with the $z_{\rm G2}=2.714$ scenario providing the most self-consistent explanation of the observed properties.

\begin{figure}
\centering
\includegraphics[scale=0.4]{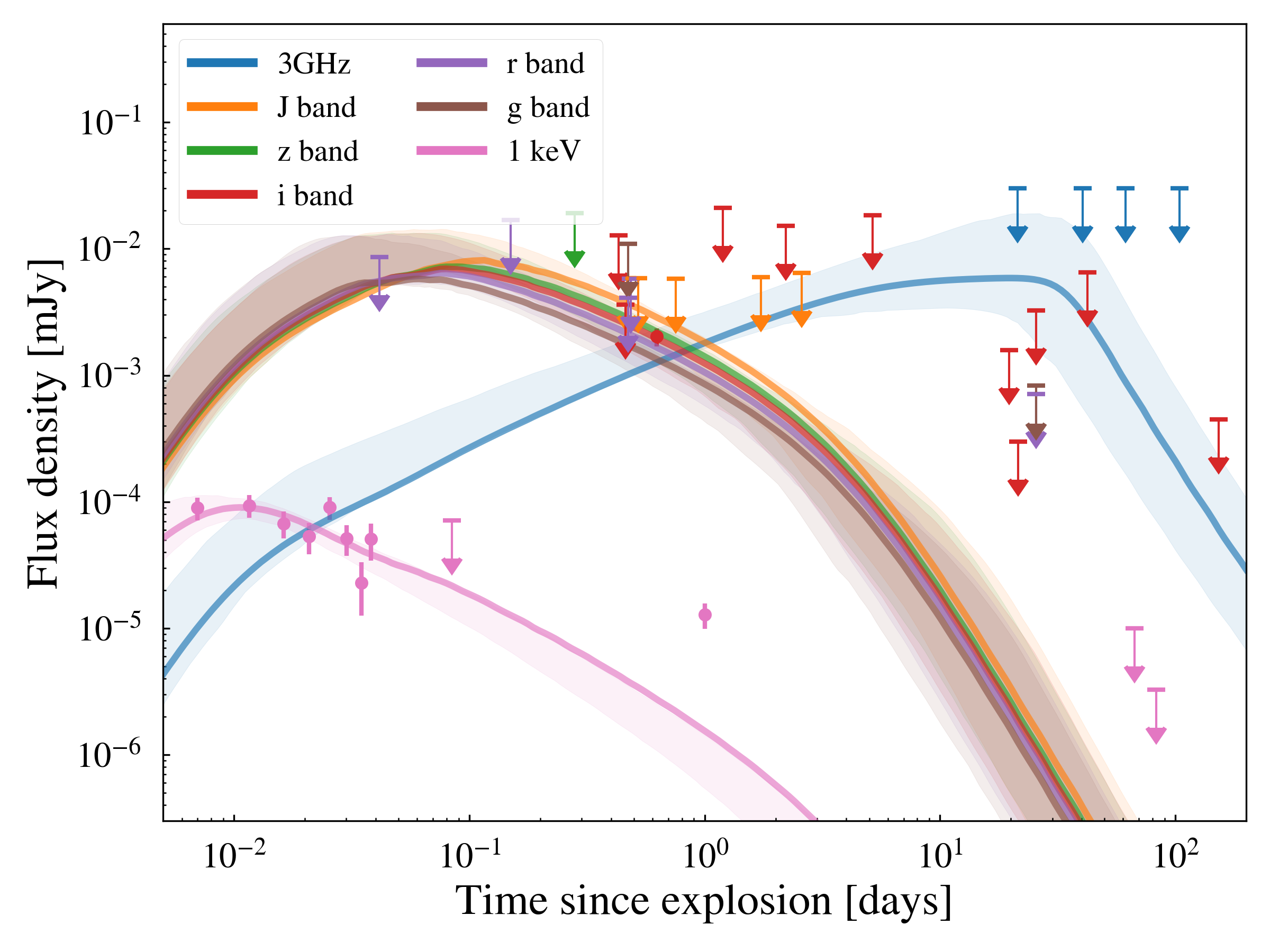}
\vspace{-0.3cm}
\caption{Multi-wavelength best fit of EP250905a, using \textsc{redback} \citet{Sarin2024}, at an assumed redshift of $z_{\rm{G2}}=2.714$. The afterglow model used was the Gaussian redback model \citep{Lamb2018b}, including synchrotron self-absorption effects \citep{Lamb2019a}.
Each color depicts a different energy band, where the solid lines and the shaded regions indicate the maximum likelihood model and the 90\% confidence interval, respectively, for each energy band.
}
\label{fig:modeling}
\end{figure}

To test the hypothesis of the afterglow nature of EP250905a at redshift $z_{\rm G2}=2.714$, we modeled the multi-wavelength data using \textsc{redback} \citep{Sarin2024}. We compared the radio, optical, NIR, and X-ray data of EP250905a with those estimated via the \texttt{afterglow\_models.gaussian\_redback} structured jet model \citep{Lamb2018b}, which assumes a Gaussian-shaped jet structure, constant ISM, and utilizes \texttt{NESSAI} \citep{Williams2024} as the sampler with a Gaussian likelihood.
The model considers as free parameters the observers' line-of-sight angle ($\theta_{\rm{observer}}$), isotropic equivalent kinetic energy ($\log E_{\rm{K,iso}}$), jet core half opening angle ($\theta_{\rm{core}}$), angular extent of structured jet ($\theta_{\rm{edge}}$), ambient medium number density ($\log n_{\rm{ISM}}$), and the initial bulk Lorentz factor ($\Gamma_0$). Due to the low number of data points and especially data points where the source was detected, we fix some of the afterglow fit parameters such as the synchrotron participation fraction ($\xi_N=1$), the fraction of energy in electrons ($\log\epsilon_e$) to $-0.66$ and that in the magnetic field ($\log\epsilon_B$) to $-2.0$ (both values are after \citep[][]{Santana2014}) and the electron energy distribution index ($p=2.6$ from our optical-to-X-ray SED fitting). Table~\ref{tab:prior_posterior} depicts the prior distribution (and posteriors) considered in our modeling.

Figure~\ref{fig:modeling} shows the multi-wavelength light curve of EP250905a, and the best-fit (where solid lines and shaded regions indicate the maximum likelihood model and the 90\% confidence interval, respectively) using the \texttt{afterglow\_models.gaussian\_redback} model. The output Bayesian evidence is $\log(Z)=68.19\pm0.13$. Overall, the model suggests an out-of-jet-core viewing angle, with a slightly off-axis angle of $\theta_{\rm{observer}}=5.7_{-1.1}^{+1.7}$~deg.~and a jet core angle of $\theta_{\rm{core}}=3.4\pm 1.1$~deg.~(uncertainties at the 68\% confidence level), with a high constant density ISM. Comparing both parameters and considering their uncertainties, an on-axis scenario is also possible. Furthermore, the Gaussian structured jet wings extend to $\theta_{\rm{edge}}=6.9_{-2.3}^{+2.9}$~deg. After the initial decay from the EP-WXT detection (see Fig.~\ref{fig:x_flux}; we emphasize that the prompt EP-WXT emission is not modeled here, only the afterglow emission is modeled), the multi-wavelength data are compatible with a mildly off-axis afterglow emission with an initial Lorentz factor of $\Gamma_0\sim350$. We should interpret the above parameters with caution due to the limited number of constraints available in the modeling. Nevertheless, they provide a general indication of the afterglow-like parameters associated with the event.

\subsection{Gravitational Lensing posibility}

The gravitational lensing effect in transients has been robustly observed in SNe \citep{Rodney2021,Paynter2021,Magee2023}, while for GRBs, most cases remain lensing candidates \citep{Liao2022,Levan2025b}. 
The configuration of EP250905a is consistent with gravitational lensing, as it lies in close angular proximity to a massive foreground galaxy G1, providing the necessary potential to deflect and magnify background light. Its high inferred luminosity, together with this alignment, might naturally be explained by lensing-induced magnification rather than intrinsic brightness alone.
To explore this possibility, we adopt a singular isothermal sphere (SIS) mass profile\footnote{In this model, the total mass distribution of the lens is approximated as a spherically symmetric system with a constant velocity dispersion.}, which predicts an Einstein radius of $\theta_E = 4\pi \left(\sigma/c\right)^2 \left(D_{\rm{ds}}/D_{\rm{s}}\right)$ \citep{Schneider1992,Narayan1996},
where $\sigma$ is the velocity dispersion of the lens, $D_{\rm{ds}}$ and $D_{\rm{s}}$ are the angular diameter distances between the lens and source, and to the source, respectively, and $c$ is the speed of light. To estimate the velocity dispersion of galaxy G1, we fit its GTC/OSIRIS+ spectrum using the package \texttt{BAGPIPES} \citep{Carnall2018,Carnall2019} assuming an exponentially declining star formation history, obtaining a posterior value of $\sigma=300\pm50\,\mathrm{km\,s^{-1}}$.
To derive the Einstein radius and the magnification we implement a Markov Chain Monte Carlo (MCMC) method, assuming uniform priors for $\sigma$ and angular offset of $\mathcal{U}{\sim}\left[250;350\right]$~km~s$^{-1}$ and $\mathcal{U}{\sim}\left[2.2;3.0\right]$~arcsec, respectively.
We derive an Einstein radius of $\theta_E=1.9\arcsec\pm0.4$\arcsec~(at 68\% confidence level). Given the observed angular separation of ${\approx}2.6\arcsec$, the source lies outside the Einstein radius, corresponding to the single-image regime of the SIS model, and indicating moderate amplification without the formation of multiple images. In this configuration, the expected magnification is $\mu=3.9_{-1.4}^{+5.7}$ (at 68\% confidence level), showing a large uncertainty. Assuming the lower magnification limit of $\mu\approx2.5$, the intrinsic optical counterpart of EP250905a would have a magnitude of $m_i\approx24.2$~mag or fainter.
Therefore, although EP250905a may be a weak gravitational lensing source, the current data do not allow us to robustly establish this interpretation.

\section{Conclusions}\label{sec:conclusion}

We have presented a comprehensive multi-wavelength analysis of the fast X-ray transient EP250905a, combining X-ray, optical, NIR, and radio observations obtained from hours to months after the trigger.
The optical counterpart of EP250905a is identified in early imaging obtained with GTC/OSIRIS+ in the $i$-band with $m_i=23.23\pm0.18$~AB~mag. However, it remains undetected in subsequent observations across optical, NIR, and radio wavelengths. The redshift of EP250905a was not directly measured; instead, two galaxies (G1 and G2) located close to the transient position were identified as potential host candidates. Their redshifts were determined through the identification of spectral absorption ($z_{\rm{G1}}=0.374$) and emission ($z_{\rm{G2}}=2.714$) lines. Based on the lower chance-alignment probability, we find that the G2 galaxy is the most likely host galaxy, implying a fiducial redshift of $z=2.714$ for EP250905a.

The observational properties of the transient are best explained by an afterglow scenario, with modeling favoring a mildly off-axis structured jet at redshift $z_{\rm{G2}}=2.714$, which provides a self-consistent interpretation of the broadband data. Alternative scenarios at low redshift ($z_{\rm{G1}}=0.374$) are disfavored, as they require unusually faint emission across multiple wavelengths and are less consistent with the observed constraints.

The close angular proximity of EP250905a to the foreground galaxy G1 makes it an intriguing candidate for gravitational lensing. Under a singular isothermal sphere model, we estimate an Einstein radius of $\theta_E{\approx}1$\farcs9, placing the source outside the strong-lensing regime and implying a moderate magnification of $\mu{\approx}3.9$. However, the current data do not allow us to robustly confirm the lensing nature of the transient.

Overall, EP250905a represents a compelling moderate-redshift transient, potentially offering insight into both relativistic jet physics and the role of gravitational lensing in shaping the observed properties of distant explosive events. Future early imaging and spectroscopic observations of new transients will be crucial to constrain the lensing configuration and definitively establish the nature of similar sources.

\begin{acknowledgements}

J.Q.V., P.G.J., J.N.D.D., J.S.S., M.E.R., and A.P.C.H. are supported by the European Union (ERC, Starstruck, 101095973, PI Jonker). Views and opinions expressed are, however, those of the author(s) only and do not necessarily reflect those of the European Union or the European Research Council Executive Agency. Neither the European Union nor the granting authority can be held responsible for them. 

J.Q.V. additionally acknowledges support by the IAU-Gruber foundation.

DMS acknowledges support through the Ramon y Cajal grant RYC2023-044941, funded by MCIU/AEI/10.13039/501100011033 and FSE+. MAPT acknowledges support by the Spanish Ministry of Science via the Plan de Generacion de conocimiento PID2021-124879NB-I00.
F.E.B. acknowledges support from ANID-Chile BASAL CATA FB210003 and FONDECYT Regular 1241005.

A.S. acknowledges financial support from the Centre national d’études spatiales (CNES), France (ROR: \url{https://ror.org/04h1h0y33}) within the framework of the SVOM mission.

This work is based on the data obtained with Einstein Probe, a space mission supported by the Strategic Priority Program on Space Science of Chinese Academy of Sciences, in collaboration with the European Space Agency, the Max-Planck-Institute for extraterrestrial Physics (Germany), and the Centre National d'Études Spatiales (France).

We thank the Fermi GBM Team for providing assistance in producing the Fermi-GBM upper limits shown in this work.

Data for this paper have in part been obtained under the International Time Programme of the CCI (International Scientific Committee of the Observatorios de Canarias of the IAC) with the NOT and GTC operated on the island of La Palma by the Roque de los Muchachos.

The MeerKAT telescope is operated by the South African Radio Astronomy Observatory, which is a facility of the National Research Foundation, an agency of the Department of Science and Innovation. This work has made use of the “MPIfR S-band receiver system” designed, constructed and maintained by funding of the MPI für Radioastronomy and the Max-Planck-Society.

\end{acknowledgements}

\bibliographystyle{bibtex/aa}
\bibliography{Quirola-Vasquez}


\begin{appendix}

\section{EP-WXT spectra}

\begin{figure}[h!]
\centering
\includegraphics[scale=0.76]{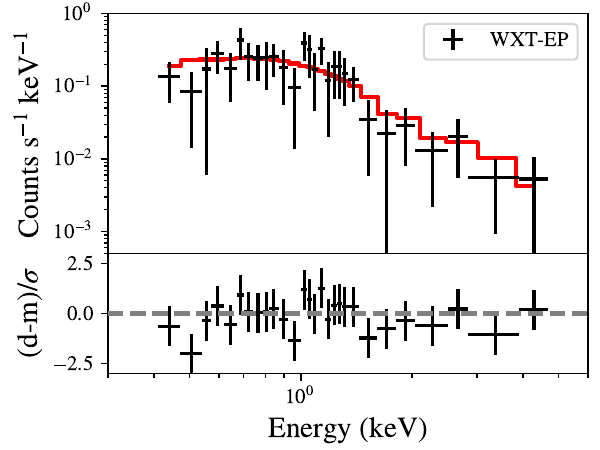}
\caption{{\it Top panel:} The average EP-WXT spectrum (0.5--4 keV) with the best-fit power law model ($\Gamma_{\rm{WXT}}=1.7\pm0.4$) affected by Galactic extinction ($N_{\rm{H,Gal}}=7\times10^{20}$~cm$^{-2}$) are shown in black and red, respectively. {\it Bottom panel:} The data, minus the best fit model, divided by the error in each data point, is shown. No significant deviations with respect to the best-fit model are present. }
\label{fig:SP-wxt}
\end{figure}

\section{EP-FXT spectra}

\begin{figure}[h!]
\centering
\includegraphics[scale=0.67]{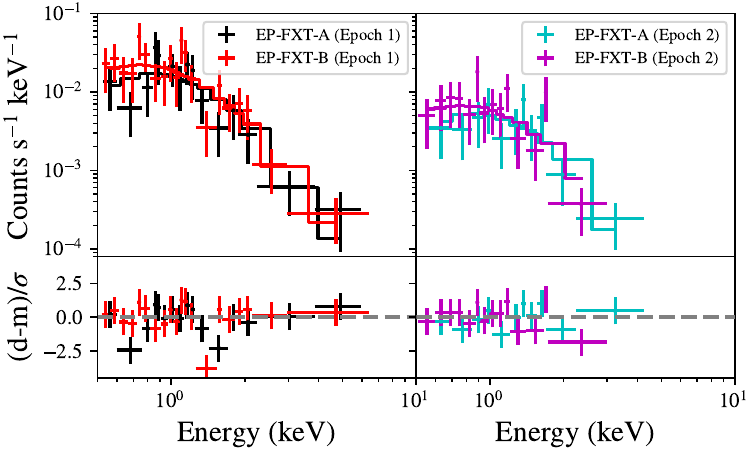}
\caption{{\it Top panels:} The average EP-FXT spectra (0.5--10~keV) for telescope units A and B separately for the first (left) and second (right) follow-up epochs taken on 2025-09-05 11:13:53 (i.e., $\Delta t\approx$0.02~days) and 2025-09-06 10:22:03 UT (i.e., $\Delta t\approx$1.0~days), respectively. The best-fit power law model ($\Gamma_{\rm{FXT}}^{\rm{E1}}=2.4\pm0.3$ and $\Gamma_{\rm{FXT}}^{\rm{E2}}=2.3\pm0.5$) affected by Galactic extinction is shown. 
{\it Bottom panels:} The data, minus the best fit model, divided by the error in each data point, are shown. No significant deviations with respect to the best-fit model are present. }
\label{fig:SP-fxt}
\end{figure}

\newpage

\section{High energy spectral energy distribution}

\begin{figure}[htbp]
\centering
\includegraphics[scale=0.8]{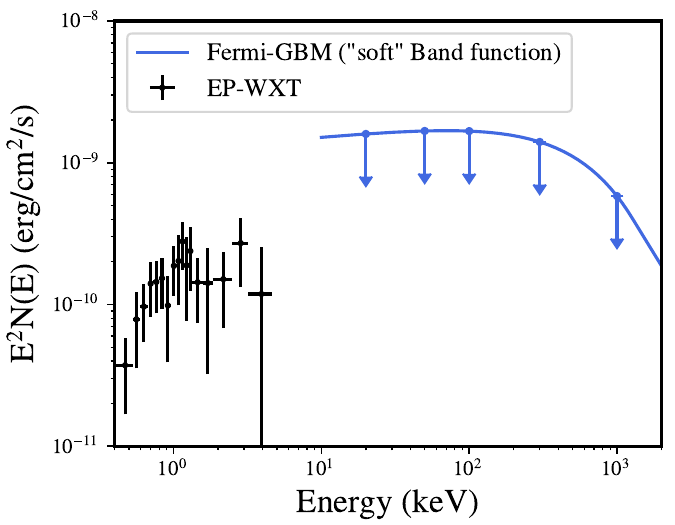}
\caption{The spectral energy distribution from soft X-ray to gamma-rays, showing the average EP-WXT discovery spectrum and the \emph{Fermi}-GBM ''soft'' band function \citep[i.e., $E_{\rm peak}=70$~keV, $\alpha=-1.9$, and $\beta=-3.7$;][]{Band1993} upper limit normalized to $7.06\times10^{-9}$~erg~cm$^{-2}$~s$^{-1}$.}
\label{fig:high_sed}
\end{figure}

\section{Afterglow modeling posterior distribution}
 
\begin{table}[htbp]
    \caption{Priors and marginalized posteriors (median, 16th/84th percentiles) for the  \textsc{redback} structured Gaussian-jet fit of EP250905.}
     \centering
     \begin{tabular}{lcc}
    \hline\hline
    Parameter & Prior & Posterior \\ \hline
    $z$ & $2.714$ (fixed) & --- \\
    $\theta_{\rm observer}$ [rad] & $\mathrm{Sine}(0,\pi/2)$ & $0.10_{-0.02}^{+0.03}$ \\
    $\log_{10}(E_{0}/{\rm erg})$ & $\mathcal{U}(44,54)$ & $52.02_{-0.14}^{+0.13}$ \\
    $\theta_{\rm core}$ [rad] & $\mathcal{U}(0.01,0.1)$ & $0.06\pm0.02$ \\
    $\theta_{\rm edge}$ [rad] & $\mathcal{U}(0.05,0.5)$ & $0.12_{-0.04}^{+0.05}$ \\
    $\log_{10}(n_{0}/{\rm cm^{-3}})$ & $\mathcal{U}(-5,2)$ & $-0.07\pm0.27$ \\
    $p$ & 2.6 (fixed) & -- \\
    $\log_{10}\epsilon_{e}$ & $-$0.66 (fixed) & -- \\
    $\log_{10}\epsilon_{B}$ & $-$2.0 (fixed) & -- \\
    $\Gamma_{0}$ & $\mathcal{U}(40,400)$ & $345.8_{-68.35}^{+38.51}$ \\
    $\xi_{N}$ & 1.0 (fixed) & -- \\
    \hline
     \end{tabular}
     \tablefoot{Parameters are the viewing  angle $\theta_{\rm observer}$, isotropic kinetic energy $E_{0}$, jet core and truncation angles $\theta_{\rm c}$ and $\theta_{\rm j}$, circumburst density $n_{0}$, electron  power-law index $p$, electron and magnetic energy fractions $\epsilon_{e}$ and  $\epsilon_{B}$, initial Lorentz factor $\Gamma_{0}$, and fraction of accelerated  electrons $\xi_{N}$.}
     \label{tab:prior_posterior}
 \end{table}

\begin{figure}[htbp]
\centering
\includegraphics[scale=0.28]{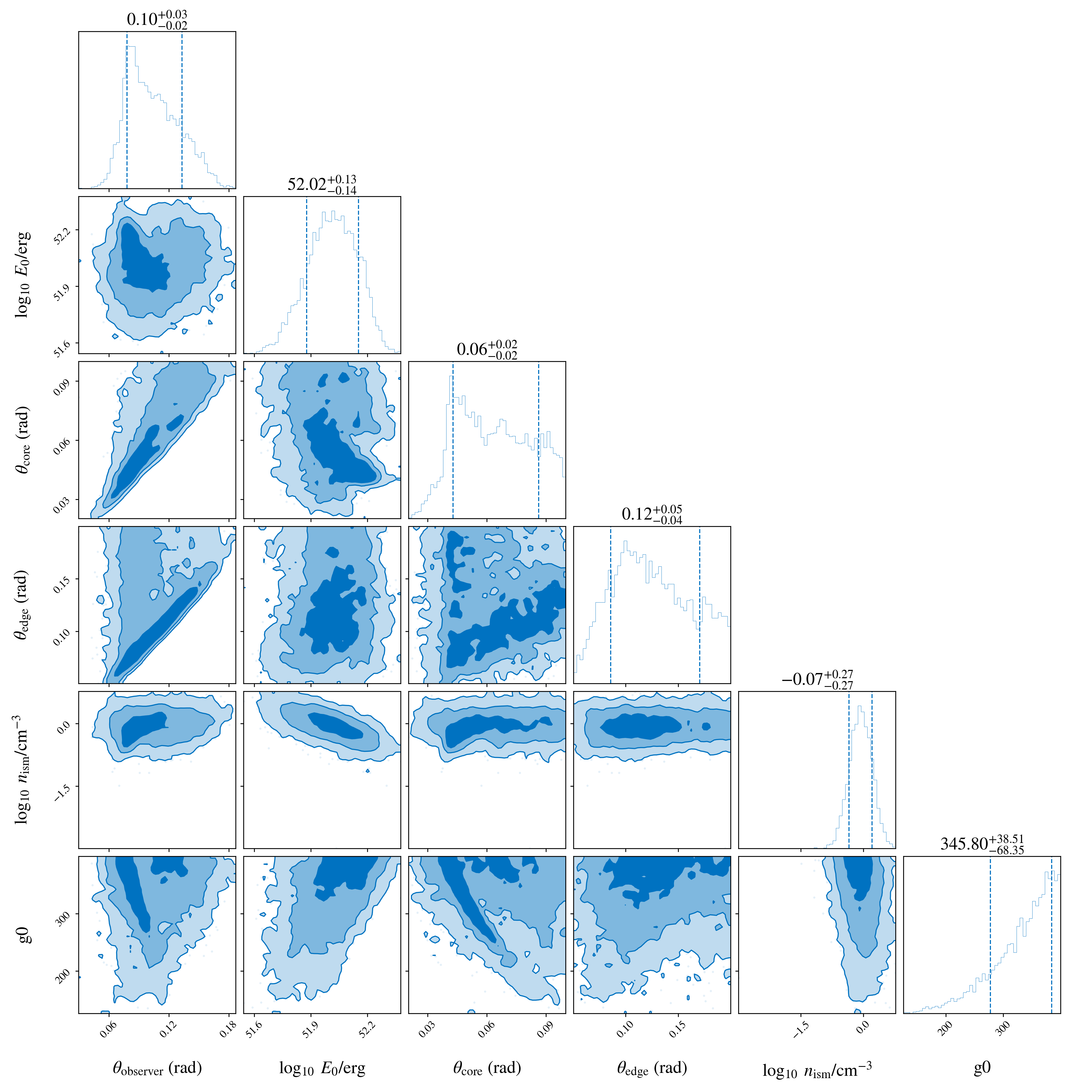}
\caption{The posterior distribution for the model parameters used to fit the afterglow in Fig~\ref{fig:modeling}.
The model fit used \texttt{NESSAI} as the sampler with a Gaussian likelihood via \textsc{redback}.}
\label{fig:corner}
\end{figure}

\newpage

\section{VLT-XSHOOTER spectra}

\begin{figure}[h!]
\centering
\includegraphics[scale=0.35]{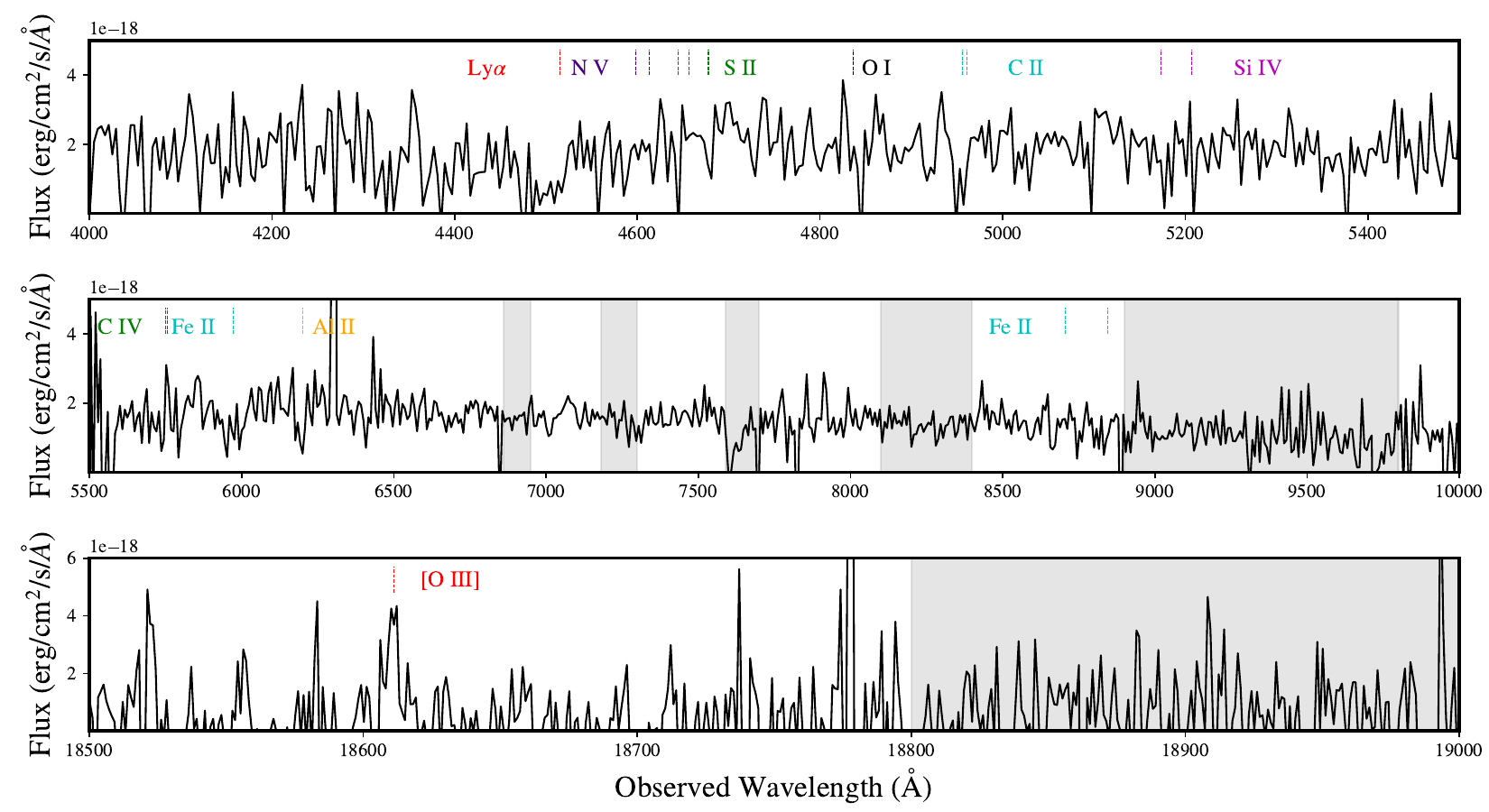}
\caption{X-shooter spectrum of the host galaxy G2 of EP250905a with the detected absorption lines from the host system (color vertical lines). The gray regions denote the atmospheric absorption bands.}
\label{fig:xshooter}
\end{figure}

\newpage
\newpage

\section{Photometry and fluxes}

Tables~\ref{tab:photometry}, \ref{tab:x-rays} and \ref{tab:radio} show the optical, X-ray and radio photometry of the transient EP250905a.

\begin{table*}[h!]
    \caption{Photometry obtained with various ground-based telescopes.}
    \centering
    \scalebox{0.85}{
    \begin{tabular}{lllccccc}
    \hline\hline
    Telescope & Instrument & Date (UT) & Days since trigger & Exposure time & Filter & AB magnitude & Reference \\ 
    (1) & (2) & (3) & (4) & (5) & (6) & (7) & (8) \\ \hline
    TRT & Andor & 2025-09-05 11:56:27.3 & 0.0415 & 7$\times$180 & $R$ & $>21.70$ & 3 \\
    ALT100C & Andor & 2025-09-05 14:42:58.83 & 0.1499 & 15$\times$180 & $r$ & $>20.97$ & 3 \\
    ALT100C & Andor & 2025-09-05 17:51:09.9 & 0.2806 & 24$\times$200 & $z$ & $>20.76$ & 3 \\
    ALT100C & Andor & 2025-09-05 21:28:10.82 & 0.4313 & 25$\times$180 & $i$ & $>21.23$ & 3 \\
    Liverpool & IO:O & 2025-09-05 22:06:03 & 0.461 & 6$\times$100 & $i$ & $>22.6$ & 1 \\
    NOT & NOTCam & 2025-09-05 22:13:15.9 & 0.4678 & 15$\times$60 & $J$ & $>22.0$ & 3 \\
    Liverpool & IO:O & 2025-09-05 22:20:05 & 0.4725 & 6$\times$150 & $r$ & $>22.5$ & 2\\
    Liverpool & IO:O & 2025-09-05 22:20:05 & 0.4725 & 6$\times$150 & $g$ & $>21.5$ & 2\\
    LCO & Andor & 2025-09-05 22:34:45.86 & 0.4862 & 5$\times$300 & $r$ & $>22.13$ & 3 \\
    NOT & NOTCam & 2025-09-05 23:24:43.6 & 0.5213 & 26$\times$60 & $J$ & $>22.0$ & 3 \\
    GTC & OSIRIS+ & 2025-09-06 02:01:39 & 0.6243 & 9$\times$60 & $i$ & 23.23$\pm$0.18 & 3 \\
    NOT & NOTCam & 2025-09-06 04:58:46.0 & 0.7532 & 26$\times$60 & $J$ & $>22.01$ & 3 \\
    ALT100C & Andor & 2025-09-06 15:48:32.60 & 1.1954 & 19$\times$200 & $i$ & $>20.69$ & 3 \\
    NOT & NOTCam & 2025-09-07 04:30:28.1 & 1.7336 & 26$\times$60 & $J$ & $>21.98$ & 3 \\
    ALT100C & Andor & 2025-09-07 16:05:15.34 & 2.2070 & 18$\times$200 & $i$ & $>21.04$ & 3 \\
    NOT & NOTCam & 2025-09-08 01:01:48.4 & 2.5886 & 26$\times$60 & $J$ & $>21.96$ & 3 \\
    ALT100C & Andor & 2025-09-10 14:45:25.94 & 5.1655 & 12$\times$200 & $i$ & $>20.83$ & 3 \\
    GTC & OSIRIS+ & 2025-09-25 03:40:03 & 19.6926 & 9$\times$60 & $i$ & $>23.5$ & 3 \\
    GTC & OSIRIS+ & 2025-09-26 22:13:34 & 21.4793 & 19$\times$150 & $i$ & $>25.3$ & 3 \\
    Liverpool & IO:O & 2025-10-01 01:49:01.6 & 25.6124 & 11$\times$200 & $i$ & $>22.71$ & 3 \\
    NOT & ALFOSC & 2025-10-01 01:54:47 & 25.6216 & 3$\times$300 & $g$ & $>24.3$ & 3 \\
    NOT & ALFOSC & 2025-10-01 02:11:21 & 25.6366 & 5$\times$300 & $r$ & $>24.4$ & 3 \\
    ALT100C & Andor & 2025-10-17 18:09:27.89 & 42.2933 & 24$\times$300 & $i$ & $>21.96$ & 3 \\
    GTC & OSIRIS+ & 2026-02-04 20:35:03 & 152.4 & 9$\times$60 & $i$ & $>24.87$ & 3 \\
    \hline
    \end{tabular}
    }
    \tablefoot{\emph{Columns 1 and 2:} Telescope and instrument, respectively. \emph{Column 3 and 4:} Start date and time of the observation and mid-time in days after the X-ray trigger, respectively. \emph{Columns 5 and 6:} The exposure time and filter, respectively. \emph{Column 7:} AB magnitude or fluxes and their uncertainties. UL in the AB magnitude column stands for the 3$\sigma$ upper limit. \emph{Column 8:} Reference: (1) \citet{Bochenek2025GCN}; (2) \citet{Eyles-Ferris2025GCN}; (3) This work. The photometry was not corrected for Galactic extinction, nor for any potential host contribution.}
    \label{tab:photometry}
\end{table*}

\begin{table*}[h!]
    \caption{X-ray detections of EP250905a used in this work.}
    \centering
    \scalebox{0.85}{
    \begin{tabular}{lllcccccc}
    \hline\hline
        Telescope & Instrument & Date (UT) & Days since trigger & Exp. time (ks) & Band (keV) & $\Gamma$ & Flux (erg~s$^{-1}$~cm$^{-2}$) & Reference \\
        (1) & (2) & (3) & (4) & (5) & (6) & (7) & (8) & (9) \\
        \hline
         Einstein Probe & WXT &  2025-09-05 11:07:08 & Trigger & -- & 0.5--4.0 & $1.7\pm0.4$ & $\left(4.2\pm1.1\right)\times10^{-10}$ & This work \\
         Einstein Probe & FXT &  2025-09-05 11:13:53 & 0.02 & 3.0 & 0.5--10.0 & $2.4\pm0.3$ & $\left(3.4_{-0.6}^{+0.8}\right)\times10^{-13}$ & This work \\
         Einstein Probe & FXT &  2025-09-06 10:22:03 & 1.0 & 6.0 & 0.5--10.0 & $2.3\pm0.5$ & $\left(1.4_{-0.3}^{+0.5}\right)\times10^{-13}$ & This work \\
         Einstein Probe & FXT &  2025-11-11 09:10:26 & 66.92 & 1.2 & 0.5--10.0 & $2.3$ (fixed) & $<3\times10^{-13}$ & This work \\
         Einstein Probe & FXT &  2025-11-27 12:09:4 & 83.09 & 9.3 & 0.5--10.0 & $2.3$ (fixed) & $<9\times10^{-14}$ & This work \\
        \emph{Swift} & XRT & 2025-09-05 12:58:3 & 0.084 & 1.2 & 0.3--10 & $2.3$ (fixed) & $<7.9\times10^{-13}$ & This work \\
    \hline
    \end{tabular}
    }
    \tablefoot{\emph{Columns 1 and 2:} Telescope and instrument per observation, respectively. \emph{Column 3 and 4:} start date of the observation and days after the X-ray trigger, respectively. \emph{Columns 5 and 6:} Exposure time and energy band per observation, respectively. \emph{Column 7:} photon index computed or assumed. \emph{Column 8:} Unabsorbed flux. \emph{Column 9:} Reference.}
    \label{tab:x-rays}
\end{table*}

\begin{table*}[h!]
    \caption{Radio observations of EP250905a used in this work.}
    \centering
    \resizebox{0.8\textwidth}{!}{
    \begin{tabular}{llcccc}
        \hline\hline
        Telescope & Start date (UT) & Since trigger (d) & Frequency (GHz) & Flux density ($\mu$Jy) & Reference \\ 
        (1) & (2) & (3) & (4) & (5) & (6) \\ \hline
        MeerKAT & 2025-09-27 & 21.4167  & 3 GHz & $<$30 & This work \\
        MeerKAT & 2025-10-16 & 40.3667 &  3 GHz & $<$30 & This work \\
        MeerKAT & 2025-11-06 & 61.3467 &  3 GHz & $<$30 & This work \\
        MeerKAT & 2025-12-19 & 104.2367 &  3 GHz & $<$30 & This work \\
        \hline
    \end{tabular}
    }
    \tablefoot{\emph{Column 1:} telescope per observation. \emph{Columns 2 and 3:} start day of the observation and the day since the X-ray trigger, respectively. \emph{Columns 4, 5 and 6:} frequency, flux density per observation (mJy units), and references, respectively.}
    \label{tab:radio}
\end{table*}

\end{appendix}
\end{document}